\documentclass[11pt]{article}

\usepackage{geometry}
\usepackage{amsmath,amssymb}
\usepackage{changepage}
\usepackage{placeins}
\usepackage{commath}
\usepackage[capitalize]{cleveref}
\usepackage{bm}
\usepackage{ragged2e}
\usepackage[d]{esvect}
\newcommand{\q}[1]{``#1''}
\usepackage{xcolor}

\usepackage{authblk}
\usepackage{graphicx,epstopdf}
\usepackage{url}
\usepackage[numbers,sort&compress]{natbib}
\usepackage{makecell}

\usepackage{xparse}
\usepackage{float}

\usepackage[title]{appendix}

\newcommand{\MPM}{{\bm{\varphi}_c}}
\newcommand{\PM}{{\bm{\varphi}}}
\newcommand{\PSI}{{\bm{\psi}}}
\newcommand{\DATA}{{\bm{d}}}
\newcommand{\INPUT}{{\bm{x}}}
\newcommand{\GIVEN}{\hspace{0.05em}  | \hspace{0.15em}}
\newcommand{\CMODEL}{\mathcal{C}}
\newcommand{\MODEL}{\mathcal{M}}
\newcommand{\REAL}{ \mathbb{R} }

\newcommand{\COMMA}{ \; ,}
\newcommand{\PERIOD}{ \; .}
\newcommand{\NORMAL}{ \mathcal{N} }

\newcommand{\VDATA}{\vv{\DATA}} 

\newcommand{\M}{\mathcal{M}} 
\newcommand{\MU}{\mathcal{U}} 

\newcommand{\vx}{\vec{x}}
\newcommand{\vy}{\vec{d}} 
\newcommand{\vvy}{\vv{\vy}} 
\newcommand{\C}{\vec{\Sigma}} 

\newcommand{\p}{\varphi} 
\newcommand{\vp}{\vec{\varphi}} 

\newcommand{\hp}{\psi} 
\newcommand{\vhp}{\vec{\psi}} 

\newcommand{\N}{\mathcal{N}}
\newcommand{\U}{\mathcal{U}} 

\newcommand{\R}{\mathbb{R}}

\DeclareDocumentCommand \pr { m o } { \IfNoValueTF{#2} {p(#1)}{p(#1 \,|\, #2)} }

\newcommand{\cnt}{\, ,} 
\newcommand{\stp}{\, .} 

\newcommand{\COLa}{black}

\newcommand{\REVa}[1]{{\color{\COLa}#1}}

\newcommand{\ga}[1]{{\color{black}#1}}

\DeclareMathOperator{\diag}{diag}

\DeclareMathOperator*{\argmax}{arg\,max}

\title{Bayesian selection for coarse-grained models of liquid water}

\author[1]{Julija Zavadlav}
\author[1]{Georgios Arampatzis}
\author[1]{Petros Koumoutsakos\thanks{petros@ethz.ch}}
\affil[1]{Computational Science and Engineering Laboratory, ETH Zurich, Clausiusstrasse 33, CH-8092, Switzerland}

\date{}

\begin{document}

\flushbottom

\maketitle

\begin{abstract}
The necessity for accurate and computationally efficient representations of water in atomistic simulations that can span biologically relevant timescales has born the necessity of coarse-grained (CG) modeling. 
Despite numerous advances, CG water models rely mostly on a-priori specified assumptions. How these assumptions affect the model accuracy, efficiency, and in particular transferability, has not been systematically investigated. 
Here we propose a data driven, comparison and selection for CG water models through a Hierarchical Bayesian framework. We examine CG water models that differ in their level of coarse-graining, structure, and number of interaction sites. We find that the importance of electrostatic interactions for the physical system under consideration is a  dominant criterion for the model selection. Multi-site models are favored, unless the effects of water in electrostatic screening are not relevant, in which case the single site model is preferred due to its computational savings. 
{\color{black} The charge distribution is found to play an important role in the multi-site model's accuracy while the flexibility of the bonds/angles may only slightly improve the models. Furthermore, we find significant variations in the computational cost of these models.} 
We present a  data informed rationale for the selection of CG water models and provide guidance for future water model designs.
\end{abstract}

\thispagestyle{empty}

\section*{Introduction}

Water, an essential constituent of life  \cite{Alberts:1997}, remains an elusive target for modeling and simulation.  
Effective coarse-grained (CG) models of liquid water must balance computational savings, by handling fewer degrees of freedom, while at the same time capturing its essential physical properties ~ \cite{Noid:2013,Shearer2018,Buslaev2017,Bell2017,Fajardo2015}. CG water models have enabled simulations exceeding micro-meters/seconds that are relevant for processes in biophysical systems that are beyond the reach of conventional atomistic molecular dynamics (MD) simulations. CG modeling entails recasting the complex and detailed atomistic model into a simpler yet accurate representation. A CG model has the ability to model key quantities of interest (QoI) when it captures the effects of the eliminated degrees of freedom (DOFs) ~ \cite{Riniker:2012,Foley:2015,Wang:2009}. The CG process requires: (i) {\it The identification of the system's optimal resolution}. Commonly, groups of atoms are described with pseudo-atoms/interaction sites and a \q{mapping} function is used to determine the relation between these sites and the atomistic coordinates. For a given system various coarse-graining levels can be employed.  For example, existing CG lipid membrane models range from representations with a single anisotropic site~ \cite{Drouffe:1991} to three sites per lipid thus differentiating between the head and the tail~ \cite{Cooke:2005,Shillcock:2005}, or by grouping three or four heavy atoms into beads thus capturing varying degrees of chemical properties~ \cite{Klein:2001,Marrink:2004,Li:2016,Essex:2011}; (ii) {\it The specification of the associated Hamiltonian}. Here, DOFs can be reduced by simplifying the form or by neglecting specific terms in the Hamiltonian. For instance, one can neglect the bond/angle vibrations and resort to rigid models~ \cite{Ciccotti2011}. 

In effective CG models, the removed DOFs are insignificant for the QoI. However, to what degree a specific DOF is negligible for a given observable is hardly ever known beforehand. Thus, the majority of the CG models are designed based on intuition or extrapolations from existing models. Additionally, the number of the removed DOFs must be large so that the diminished accuracy compared to the AT models is justified with the substantial computational gains. It is usually assumed that increasing the level of coarse-graining will decrease the model's accuracy. However, and this is perhaps a key issue, the relation between the number of employed DOF in a model and its accuracy may not be a monotonic function~ \cite{Mullinax:2009,Das:2012,Riniker:2012}. Thus, one can end up (without realizing) in a worst case scenario, where the constructed model is redundant, i.e., better accuracy can be achieved with fewer DOFs (computationally less demanding model). For example, the two-site and four-site models of n-hexane molecule perform reasonably well, while a very similar three-site model fails~ \cite{Das:2012}. A number of works have addressed the systematic selection of CG models in bio-molecular systems ~ \cite{Sinitskiy:2012,Arkhipov:2008,Rudzinski:2014,Zhang:2008,reith,Voth:2008}.

The challenge of striking the optimal balance between accuracy and computational cost is crucial for CG models of water.  At the same time, obtaining water-water interactions consumes the majority of the computational effort. Thus, many CG models of water were developed. These models differ in the coarse-graining resolution level, i.e., the mapping, which ranges from $1$ to $11$ water molecules per CG bead~ \cite{Noid:2013,Hadley:2012}. 
 Models also differ in the employed Hamiltonian. For CG models where one bead represents one water molecule (1-to-1 mapping), the Hamiltonian is either derived from the atomistic simulations~ \cite{Chaimovich:2009,Wang:2009} or parametrized based on analytic potentials ranging from a simple Lennard-Jones (LJ) to potentials incorporating tetrahedral ordering, dipole moment, and orientation-dependent hydrogen bonding interactions~ \cite{Izvekov:2005,Molinero:2009,Jagla:1999,Hynninen:2012}. On a higher coarse-graining level, it was soon realized that chargeless models, such as the standard MARTINI model~ \cite{Marrink:2013,adaptiveprotein}, introduce unphysical features when applied to interfaces, such as an interface between water and a lipid membrane~ \cite{pmw,bmw,adaptivepolarizable}. Thus, new CG models were developed which treat the electrostatics explicitly. In the PCGS model ($3$-to-$1$)~ \cite{Ha-Duong:2009}, the CG beads carry induced dipoles, in the polarizable MARTINI model ($4$-to-$3$)~ \cite{pmw} the electrostatic is modeled analog to the Drude oscillator, in the BMW model ($4$-to-$3$)~ \cite{bmw} the CG representation resembles a rigid water molecule with a fixed dipole and quadrupole moment, while the GROMOS CG model ($5$-to-$2$)~ \cite{GunsterenFGwater} introduces explicit charges with a fluctuating dipole. Note that in these models the extra interaction sites have no relation to the physical system making the intuitive construction of the model even more difficult.

Thus far, studies reporting the effects of the choices made in the coarse-graining level and model structure are relatively few. For water, the mapping was investigated by Hadley et al.~ \cite{Hadley:2010}, where the investigated CG models were single-site models and the Hamiltonian was parameterized to reproduce the structural properties of water. The mapping $4$-to-$1$ was found to give the optimal balance between efficiency and accuracy. However, by comparing the properties of the available water models it is hard to extract any physics as the models were developed to reproduce different properties. Furthermore, one should avoid artificially constructed scoring functions that could be biased but rather perform model selection based on rigorous mathematical foundation. In this respect, the Bayesian statistical framework can serve as a powerful tool which has become a popular technique to refine, guide and critically assess the MD models~ \cite{Angelikopoulos:2012,Angelikopoulos:2013,Kulakova:2017,Jacobson:2014,Rizzi:2013,Farrell:2015}. 

\begin{figure*}[h]
\centering
\includegraphics[width = 0.8\textwidth]{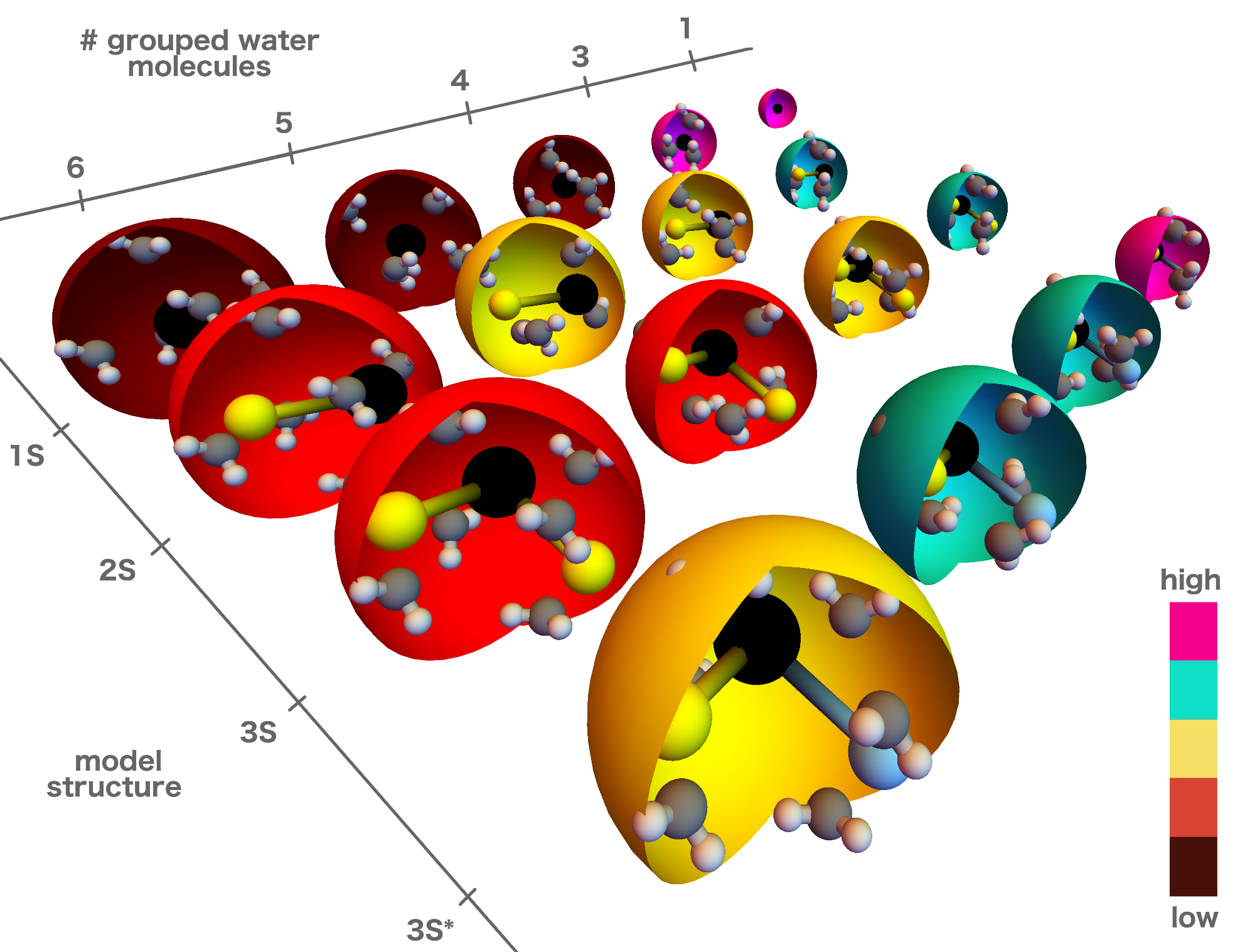}
\caption{Schematic representation of the investigated models with rigid geometry. We consider several  levels of coarse-graining and model structures. The number of grouped water molecules shown ranges from $3$ ($1$ for the {\it 1S} model) to $6$. The 4 model structures, i.e., the {\it 1S, 2S, 3S}, and {\it 3S*} are explained in the text. The spheres are color-coded according to the model's evidence rank (pink, green, yellow, red, dark red denote high to low model evidences, respectively).  }
\label{model}
\end{figure*}

In this work, we employ the Bayesian statistical framework to critically assess many CG water models (see \cref{model}).
We investigate the biologically relevant CG resolution levels, i.e., mappings, where the number of grouped water molecules ranges from $1$ to $6$. At each resolution, multiple model structures are examined ranging from $1$-site to $3$-site models where we additionally investigate the rigid and flexible versions of the $2$ and $3$-site models for mapping $M=4$. Our main objective is to determine the model evidence for all {\color{black} models} and thus elucidate the impact of the mapping on the model's performance and the relevant DOFs in CG modeling of water. Furthermore, we evaluate the speed-up for each developed model which allows us to assess efficiency-accuracy trade-off. Lastly, we investigate the transferability of the water models to different thermodynamics states, i.e., to different temperatures. To this end, we employ the hierarchical Bayesian framework~ \cite{Wu:2015} that can accurately quantify the uncertainty in the parameter space for multiple QoI, i.e., different properties or the same property at different conditions. 

\section*{Methods}\label{methods}
\begin{figure*}[h]
\centering
\includegraphics[width = 0.8\textwidth]{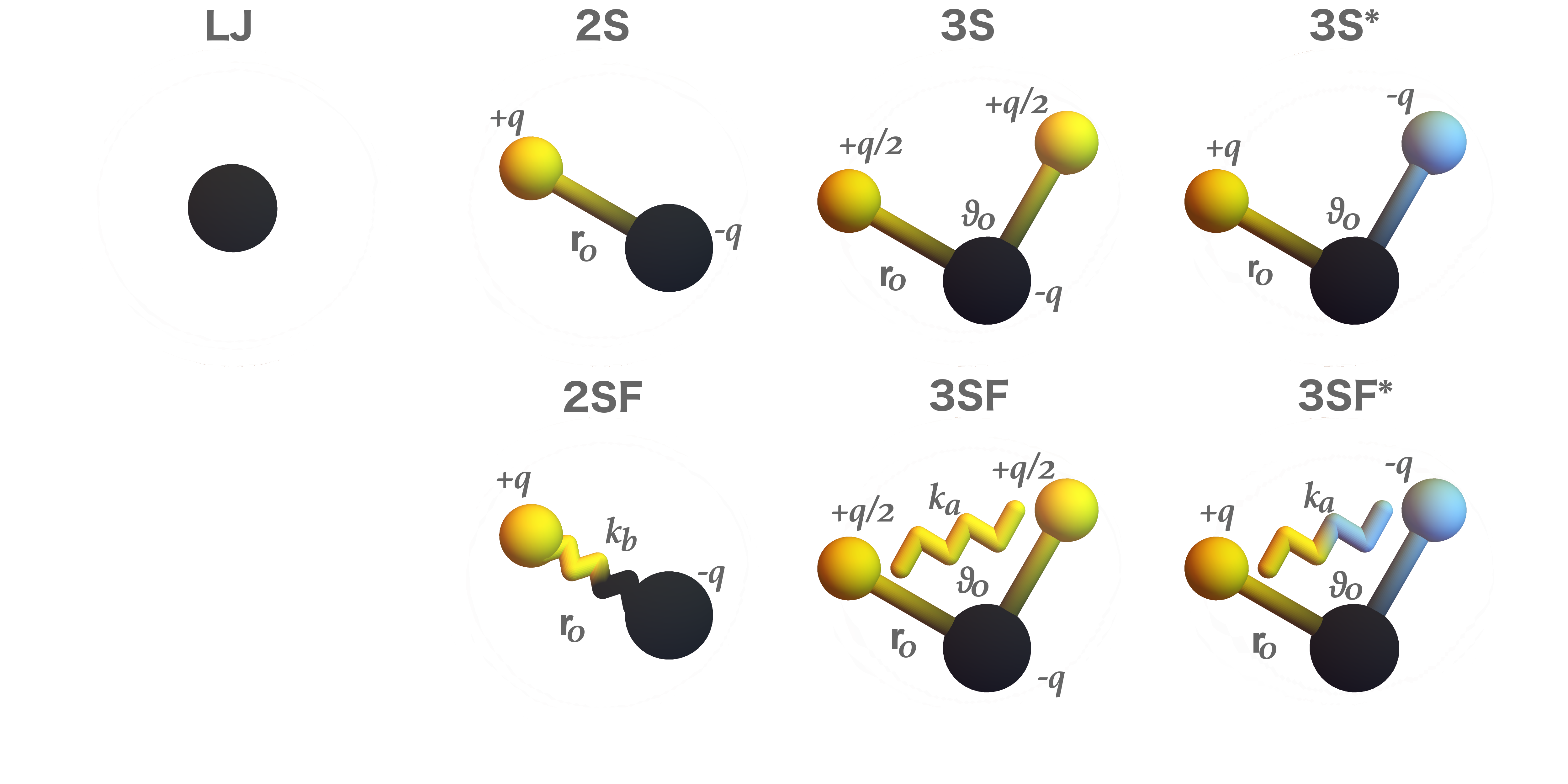}
\caption{Model structures: {\it 1S}, {\it 2S}, {\it 3S}, and {\it 3S*} from left to right. The black interaction sites interact via the LJ potential, whereas the yellow and blue sites interact only with the electrostatic interaction. In the {\it 2S}, {\it 3S} models, the black site carries a negative charge and the yellow site carries a positive charge. In the {\it 3S*} model, the black interaction site is charge neutral while the yellow and blue sites carry an opposite charge. The models in the bottom line are the flexible versions (denoted with ``{\it F}'') of the top rigid models. The rigidity or flexibility of the bonds/angles is depicted with the straight and zigzag lines, respectively. }
\label{model2}
\end{figure*}

We investigate a set of CG water models (partially shown in \cref{model}). For all models, we employ the interactions that are implemented in the standard MD packages. In the {\it 1S} model, a water cluster is modeled with a single chargeless spherical particle employing the LJ potential $U_{LJ}(r_{ij})=4\epsilon [ (\sigma/r_{ij})^{12} - (\sigma/r_{ij})^6 ]$ between particles $i$ and $j$. The model parameters are $\bm\varphi_{1S}=(\sigma, \epsilon)$. The {\it 2S} model is a two-site model, where the sites are oppositely charged ($\pm q$) and constrained to a distance $r_0$. The negatively charged (blue in \cref{model2}) site interacts additionally with the LJ potential. 
The model parameters are $\bm\varphi_{2S}=(\sigma, \epsilon, q, r_0 )$. 
In order to satisfy the net neutrality of the water cluster, the three-site model can be constructed in two ways, which we denote as {\it 3S} and  {\it 3S*} models.
The {\it 3S} model resembles a big water molecule where all three particles are charged. The central (blue) site has a charge of $-q$, and the other two sites have a charge of $+q/2$. In the {\it 3S*} model, the central site is chargeless and the other two carry a $\pm q$ charge. Both three-site models have the parameters $\bm\varphi_{3S,3S*}=(\sigma, \epsilon, q, r_0, \vartheta_0)$. For all rigid model structures, we consider four levels of resolution with the number of grouped water molecules equal to $3, 4, 5$, and $6$. For the {\it 1S} model, we additionally consider the $M=1$ mapping, while for the models with partial charges we investigate also $M=12$. The level of resolution fixes the total mass of the CG representation. 
The mass ratio between the interaction sites in the two and three-sited models is fixed to $2$ with the central particle carrying the larger mass. The electrostatic is in all cases modeled with the Coulomb's interaction $U_e(r_{ij})=q_iq_j/(4\pi \varepsilon \varepsilon_0 r_{ij})$, where we set the global dielectric screening to $\varepsilon=2.5$. 
For $M=4$, we consider also the flexible analogs of the models with charges. In the {\it 2SF} model, the two sites are interacting with a harmonic potential $U_b(r_{ij})=k_b(r_{ij}-r_0)^2$ with force constant $k_b$. Therefore, the model parameters are $\bm\varphi_{2SF}=(\sigma, \epsilon, q, r_0, k_b)$. For the flexible three-site models {\it 3SF} and  {\it 3SF*}, the angle is unconstrained and modeled with the harmonic angle potential $U_a(\vartheta_{ij})=k_a(\vartheta_{ij} - \vartheta_0)^2$ thus adding the force constant $k_a$ parameter to the parameter set, i.e., the model parameters are $\bm\varphi_{3SF,3SF*}=(\sigma, \epsilon, q, r_0, \vartheta_0, k_a)$. 

{\color{black}
We remark that the data used as target QoI is part of the modeling choice. In this work, we use experimental data of density, dielectric constant, surface tension, isothermal compressibility, and shear viscosity, i.e., mostly thermodynamic properties. These are deemed of key importance for biophysical systems. The data used and the properties of the reference coarse-grained water models are reported in~\cref{data}. The structural properties, e.g. radial distribution function or the dynamical properties, e.g. diffusion constant were not considered in this work because these properties cannot be measured experimentally for $M>1$.}
\begin{table}[h]
\centering
   \begin{tabular}{ c | c c c | c c c c}
      model & 283~K & 298~K & 323~K & \thead{MARTINI \\  \cite{Marrink:2004,pmw}} & \thead{MARTINI Pol. \\ \cite{pmw} } & \thead{BMW \\  \cite{bmw,bmw2}} & \thead{ GROMOS CG \\  \cite{GunsterenFGwater}} \\ \hline
      $\varepsilon$ [$\varepsilon_0$] &	84.0 &  78.4 & 69.9	& / 		& 75.6	& 74		& 73.7	\\
      $\gamma$ [mN/m] 	& 74.2 & 72.0 &	67.9 & 30-40	&  30.5	& 77		& 51.2	\\
      $\kappa$ [$10^{-6}$/bar] & 47.89 & 45.24	& 44.17 & 60.0	& /		& 33		& 84-138	\\
      $\eta$ [mPa s]		& 1.307 & 0.891	& 0.547 & /		& /		& 1.01-1.62 & 3.72	\\
      \end{tabular}
\caption{ {\color{black} The first three columns show the experimental data~ \cite{CRC:2004,Kell:1967} at different temperatures for density $\rho$, dielectric constant $\varepsilon$, surface tension $\gamma$, isothermal compressibility $\kappa$, and shear viscosity $\eta$ used as QoI. The remaining columns summarize the properties of reference water models at $T=298$~K used for comparison.} }
\label{data}
\end{table}
\subsection*{Uncertainty Quantification}\label{uq}

\subsubsection*{Bayesian Framework}

We consider a computational model $\CMODEL$ that depends on a set of parameters $\MPM\in\REAL^{N_\varphi}$ and a set of input variables  or conditions $\INPUT\in \REAL^{N_x}$. \ga{In the context of the current work, the computational model is the molecular dynamics solver, the model parameters correspond to the parameters of the potential and the input variables to the temperature of the system.} Moreover, we consider an observable function $F(\INPUT;\MPM)\in\REAL^N$ that represents the output of the computational model. \ga{Here, the observable function is an equilibrium property of the system, e.g., the  density.} We are interested in inferring the parameters $\MPM$ based on the a set of experimental data $\DATA = \{  d_i \,\GIVEN\, i=1,\ldots,N \}$ that correspond to the fixed input parameters of the model $\INPUT$.

In the frequentist statistics framework, the parameters of the model are obtained by optimizing a distance of the model from the data, usually the likelihood function. In the Bayesian framework, the parameters follow a conditional distribution which is given by Bayes' theorem,
\begin{equation} \label{eq:posterior}
p( \PM \GIVEN \DATA, \MODEL)  =  \frac{  p(\DATA \GIVEN \PM, \MODEL)   \; p(\PM | \MODEL)  }{   p( \DATA \GIVEN \MODEL )  }  \COMMA
\end{equation}
where $p(\DATA \GIVEN \PM, \MODEL)$ is the likelihood function, $p(\PM | \MODEL)$ is the prior probability distribution and $p(\DATA | \MODEL)$ is the model evidence. Here, $\PM$ is the vector containing the computational model parameters $\MPM$ and any other parameters needed for the definition of the likelihood or the prior density.
$\MODEL$ stands for the model under consideration and contains all the information that describes the computational and the statistical model.

The likelihood function is a measure of how likely is that the data $\DATA$ are produced by the computational model $\CMODEL$.
\REVa{Here, we make the assumption that the datum $d_i$ is a sample from the generative model
\begin{equation}
y_i = F_i( \INPUT ; \MPM ) +  \sigma_n d_i \, \varepsilon, \quad \varepsilon \sim \NORMAL(0,1) \PERIOD
\label{eq:likelihood:assumption}
\end{equation}
Namely, $y_i$ are random variables independent and normally distributed with mean equal to the observable of the model and standard deviation proportional to the data. The reason we choose this error model is because the set of experimental data $\DATA$ contains elements of different orders of magnitude, e.g., density is of order of 1 and surface tension of order $100$. With this model the error allowed by the statistical model becomes proportional to the value of the data we want to fit \cite{Moser:2011}.
The likelihood of the data $p( \DATA \GIVEN \PM )$ has the form,
\begin{equation}\label{eq:model:like}
p(\DATA | \PM ) = \NORMAL \big ( \DATA \GIVEN F( \INPUT,\MPM), \Sigma \big ) \COMMA\quad
\Sigma  =  \sigma_n^2 \diag \big( \, \DATA^2 \, \big )    \COMMA
\end{equation} 
where $\PM = ( \MPM^\top , \sigma_n)^\top$ is the parameter vector that contains the model and the error parameters.
}

The denominator of  \cref{eq:posterior} is defined as the integral of the numerator and is called the model evidence.  This quantity can be used for model selection  \cite{Beck2004} as it is discussed in the next section. Finally, the prior probability encodes all the available information on the parameters prior to observing any data. If no prior information is known for the parameters, a non informative distribution can be used, e.g. a uniform distribution. \REVa{In this work we use uniform priors, see SI for detailed information.}

\subsubsection*{Model Selection}
 Assuming we have $N_{\MODEL}$ models $\MODEL_i, \, i=1,\ldots,N_\MODEL$ that describe different computational and statistical models, we wish to choose the model that best fits the data. In Bayesian statistics, this is translated into choosing the model with the highest posterior probability,
\begin{equation}
p(\MODEL_i \GIVEN \DATA) = \frac{p(\DATA | \MODEL_i ) \, p(\MODEL_i) }{ p(\DATA) } \COMMA
\end{equation}
where $p(\MODEL_i)$ encodes any prior preference to the model $\MODEL_i$.  Assuming all models have equal prior probabilities, the posterior probability of the model depends only on the likelihood of the data. Taking the logarithm of the likelihood and using \cref{eq:posterior} we can write
\begin{equation}\label{eq:log:ev}
\begin{split}
\ln p( \DATA | \MODEL_i ) &= \int \ln p( \DATA | \MODEL_i )    \, p(  \PM |\DATA, \MODEL_i )   \dif \PM \\
&= \int \ln  \frac{  p(\DATA \GIVEN \PM, \MODEL_i)   \; p(\PM | \MODEL_i)  }{  p(  \PM |\DATA, \MODEL_i )  }    \, p(  \PM |\DATA, \MODEL_i )   \dif \PM \\
&= \mathbb{E}\left[ \ln p(\DATA \GIVEN \PM, \MODEL_i) \right] - \mathbb{E}\left[  \ln \frac{  p(\PM | \MODEL_i)  }{  p(  \PM |\DATA, \MODEL_i )  }   \right] \COMMA
\end{split}
\end{equation}
where the expectation is taken with respect to posterior probability $p(  \PM |\DATA, \MODEL_i )$.
The first term is the expected fit of the data under the posterior probability of the parameters and is a measure of how well the model fits the data. The second term is the Kullback-Leibler (KL) divergence or relative entropy of the posterior from the prior distribution and is a measure of the information gain from data $\DATA$ under the model $\MODEL_j$. The KL divergence can be seen as a measure of the distance between two probability distributions.

If one would only consider the first term of \cref{eq:log:ev} for the model selection, then the model that fits the data best would be selected. However, such an approach is prone to overfitting, i.e., choosing a too complex model, which reduces the predictive capabilities of the model. The second term serves as a penalization term. Models with posterior distributions that differ a lot from the prior, i.e., models that extract a lot of information from the data, are penalized more. Thus, model evidence can be seen as an implementation of the Ockham's razor that states that simple models (in terms of the number of parameters) that reasonably fit the data should be preferred over more complex models that provide only slight improvements to the fit. For a detailed discussion on model selection and estimators of the model evidence, we refer to Ref.  \cite{Knuth2015,Beck2010}.

\subsubsection*{Hierarchical Bayesian Framework}

We consider data structured as: $\VDATA=\{\DATA_1, \ldots, \DATA_{N_d}\}$ where $\DATA_i \in \mathbb{R}^{N_i}$  corresponds to the conditions $\INPUT_i$. For example, $\INPUT_i$ may correspond to different thermodynamic conditions under which the experimental data $\DATA_i$ are produced.  

\begin{figure*}[h!]
\centering
\includegraphics[width = 0.85\textwidth]{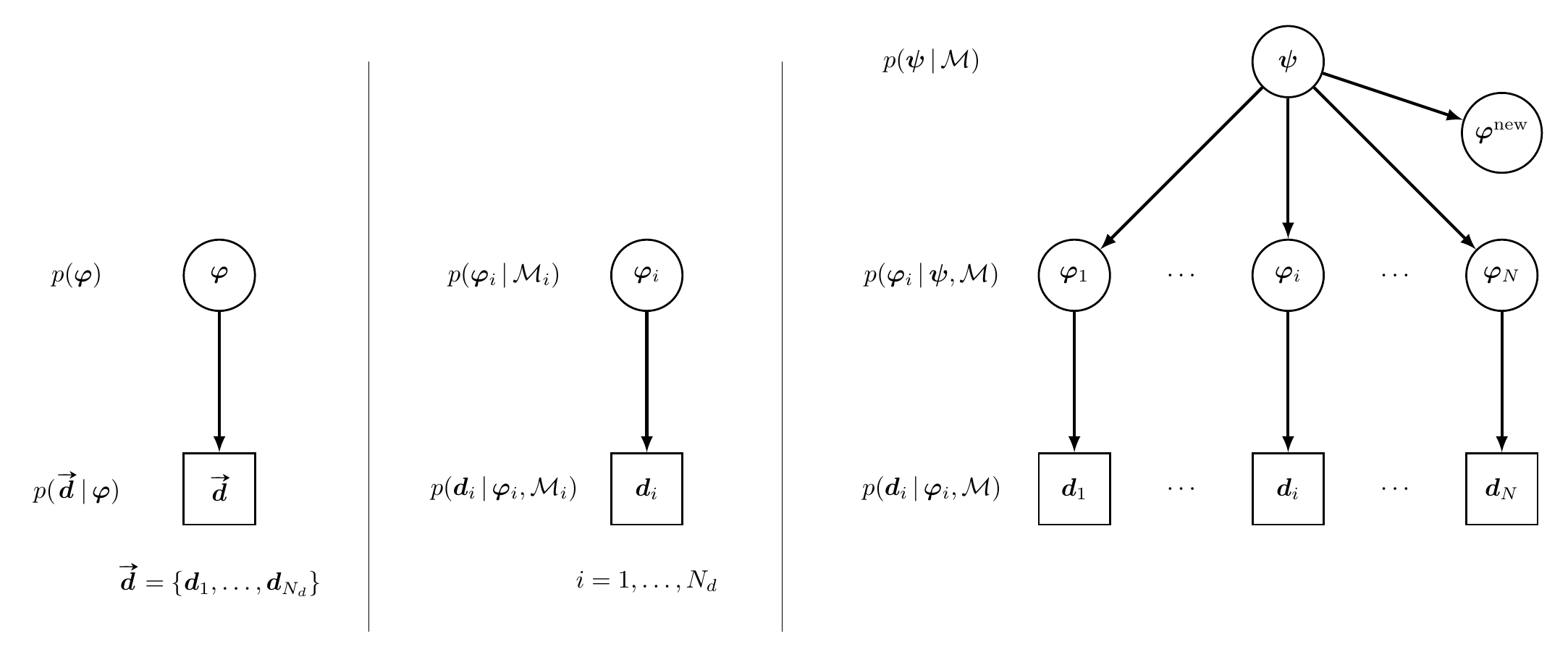}
\caption{Grouped data (left), non-hierarchical (middle) and hierarchical (right) parameter representation. In the hierarchical graph, each data set $\DATA_i$ is represented with different parameters $\PM_i$ and the parameters are connected through a hyper-parameters $\PSI$.}
\label{fig:dag}
\end{figure*}

The classical Bayesian method for inferring  the parameters of the computational model is to group all the data and estimate the probability $p(\PM \GIVEN \VDATA)$ (see \cref{fig:dag} left). However, this approach may not be suitable when  the uncertainty on $\PM$ is large due to the fact that different parameters may be suitable for different data sets. On the opposite side, individual parameters $\PM_i$ can be inferred using only the data set $\DATA_i$  (see \cref{fig:dag} middle). This approach preserves the individual information but any information that may be contained in other data sets is lost. 

Finally, a balance between retaining individual information and sharing information between different data sets can be achieved with the hierarchical Bayesian framework. In this approach, the independent models corresponding to different conditions are connected using a hyper-parameters vector $\PSI$ (see \cref{fig:dag} right). The benefits of this approach is twofold: (i) better informed individual probabilities $p(\PM_i \GIVEN \VDATA)$;  and (ii) a data informed prior $p(\PSI \GIVEN \VDATA)$ is available in case new parameters $\PM^{\textrm{new}}$ that correspond to unobserved data need to be inferred. A detailed  description of the sampling algorithm of this approach is given in Appendix.

\section*{Results}

\subsection*{Impact of mapping}
\begin{figure*}[h!]
\centering
\includegraphics[width = 0.7\textwidth]{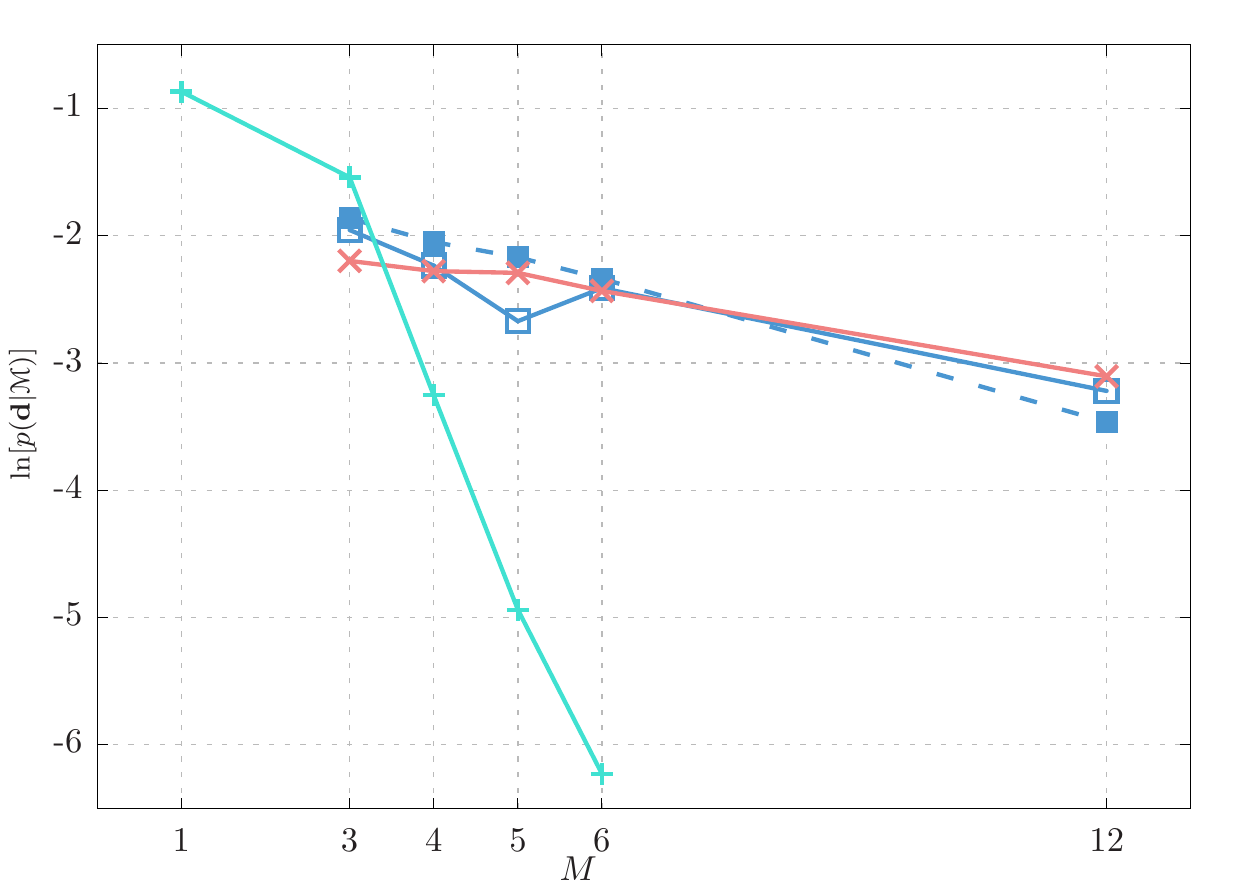}
\caption{Model evidences for the explored rigid models of liquid water: {\it 1S} ($+$), {\it 2S} ($\times$), {\it 3S} ($\boxdot$), and {\it 3S}$^*$ ($\blacksquare$). For each model, we consider different mappings $M$ ranging from $1$ to $12$, where, for example, $M=4$ means that a CG entity represents $4$ water molecules. }
\label{logE_M}
\end{figure*}

First, we examine the impact of the level of resolution on the model accuracy using density, dielectric constant, surface tension, isothermal compressibility, and shear viscosity experimental data (see SI). In \cref{logE_M} the model accuracy, as measured by the model evidence, is shown as a function of mapping $M$, which denotes the number of water molecules represented by a given CG model. 
It is usually assumed the model's performance is decreasing with the decreased resolution of the model. Indeed, for the {\it 1S} model, we observe precisely this trend. For the charged {\color{black} models}, the evidence is still overall monotonically decreasing with $M$, however, compared to the {\it 1S} model the dependency of the evidence on $M$ is much less drastic. To investigate this dependency further, we perform the UQ inference also for the charged models at $M=12$. The observed evidences are comparable to the evidence of the {\it 1S} model at $M=4$. Thus, with the models that incorporate partial charges, one can resort to models with higher mappings. According to the UQ, the best model for $M=1,3$ is the {\it 1S} model whereas for $M>3$ the charged models are superior. However, one should keep in mind that the chargeless and charged models are not comparable as the chargeless models cannot provide the same amount of information, e.g., the dielectric constant is not defined. Comparing the evidences of the {\it 2S}, {\it 3S}, and {\it 3S$^*$} models, we see that the three models rank very closely with the {\it 3S$^*$} model being somewhat better than the other two. 

\begin{figure*}[h!]
\centering
\includegraphics[width = 0.8\textwidth]{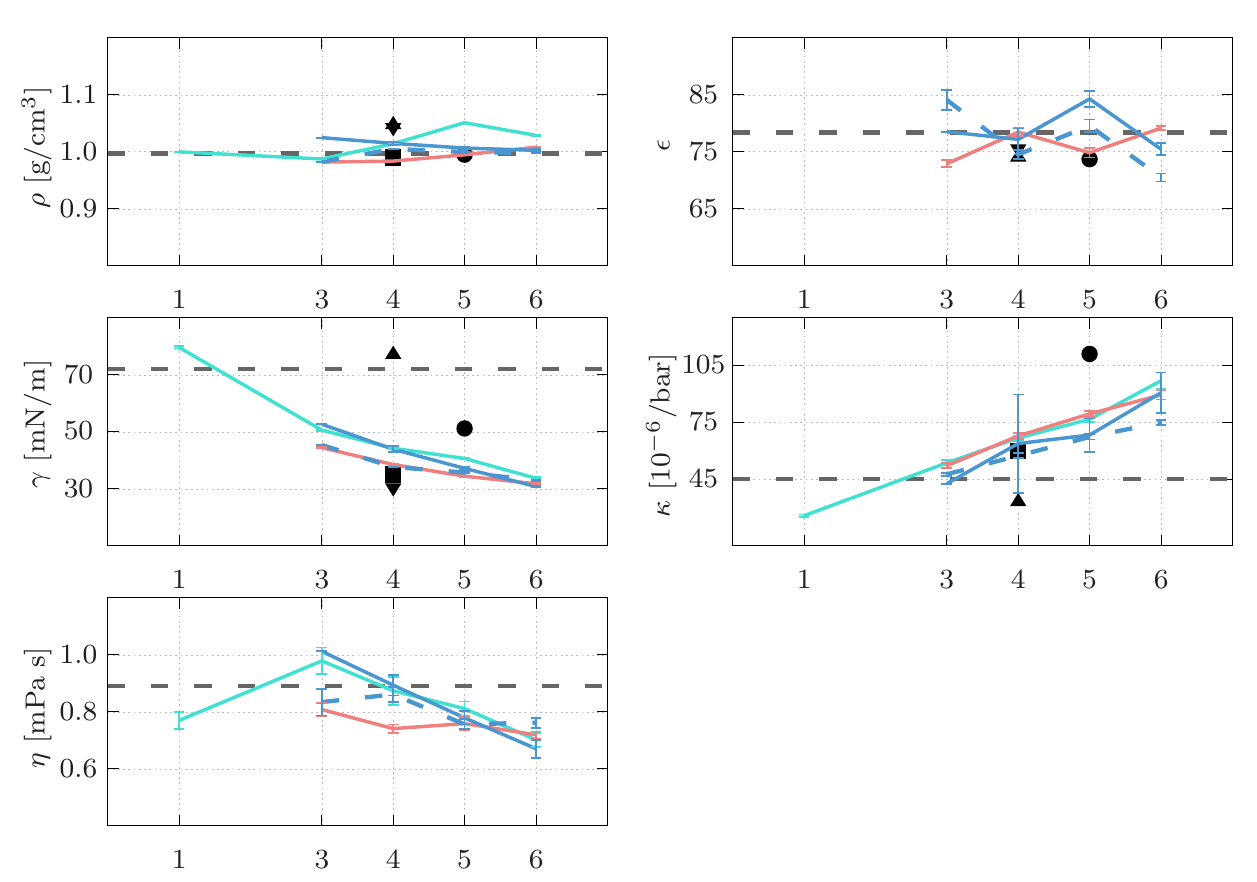}
\caption{Target QoI: density $\rho$, dielectric constant $\varepsilon$, surface tension $\gamma$, isothermal compressibility $\kappa$, and shear viscosity $\eta$ for the rigid water models {\it 1S} (green), {\it 2S} (red), {\it 3S} (blue), and {\it 3S}$^*$ ({\color{black} dashed} blue line) at different mappings $M$. The error bars denote the standard deviation of $5$ independent simulation runs with different initial conditions. The properties of the models are compared to the reported properties of the existing water models MARTINI~ \cite{Marrink:2013} ($\blacksquare$), GROMOS~ \cite{GunsterenFGwater} ($\bullet$), BMW~ \cite{bmw} ($\blacktriangle$), and polarizable MARTINI~ \cite{pmw} ($\blacktriangledown$). The experimental data is shown with the {\color{black} horizontal} dashed lines. }
\label{res}
\end{figure*}

We emphasize that the model evidence encompasses much more than a mere evaluation of the model's properties at the best parameters. Nonetheless, it is insightful to examine the target QoI and their dependency on the mapping. \cref{res} shows the density $\rho$, dielectric constant $\varepsilon$, surface tension $\gamma$, isothermal compressibility $\kappa$, and shear viscosity $\eta$ for rigid models and mappings $1$ to $6$. 
The target QoI are obtained using \REVa{the maximum a posteriori (MAP)} parameters and evaluated as a mean of $5$ independent simulation runs with different initial conditions. Note that $\varepsilon$ is not defined for the {\it 1S} model and it is excluded from the target QoI in the second UQ inference of the charged models. We observe that the $\rho$ and $\varepsilon$ are within $10\%$ of the experimental data for all mappings. On the contrary, the $\gamma, \kappa$, and $\eta$ depend very strongly on the mapping. The general trend is similar for all models, i.e., as we increase the mapping the $\gamma$ is decreasing, $\kappa$ is increasing, and $\eta$ is decreasing. This observation agrees with the general picture of coarse-graining. The more we increase the level of coarse-graining the softer are the interactions between the CG beads which correlates with increased $\kappa$ and decreased $\gamma$ and $\eta$. We observe that for some models there are no parameters $\sigma$ and $\epsilon$ {\color{black} of the LJ potential} that would fit well a certain target QoI (within the liquid state), in particular, the $\gamma$ and $\eta$. A possible solution would be to replace the LJ non-bonded interaction with another interaction, e.g, the Born-Mayer-Huggins interaction that is used in the BMW model~ \cite{bmw}. The {\it 1S} model with $M=4$ can be directly compared with the MARTINI model as the models are equal but were developed with different target QoI. With our model, we observe very similar properties as reported for the MARTINI model. Additionally, the inferred parameters with \REVa{the MAP estimates} are also very close to those of the MARTINI (see SI).

\subsection*{Rigid vs. flexible models}
For the mapping $M=4$, we examine also the three flexible models {\it 2SF}, {\it 3SF}, and {\it 3S$^*$F}. The resulting evidences for these models are listed in \cref{flex} along with the model evidences for the rigid counterparts. 
%
\begin{table*}[h!]
\begin{center}
  \begin{tabular}{ c | c ||  c | c }
    rigid model & $\log E$ & flexible model & $\log E$  \\ \hline \hline
    2S,4 & -2.28 & 2SF,4 & -2.25 \\ \hline
    3S,4 & -2.23 & 3SF,4 & -2.37 \\ \hline
    3S*,4 & -2.05 & 3S*F,4 & -2.54 \\ 
  \end{tabular}
\end{center}
\caption{Model evidences for models with mapping $M=4$. The rigid models are compared with the flexible counterparts.}
\label{flex}
\end{table*}
The physical motivation behind the flexible models is that they encompass the fluctuations in the dipole moment of the water cluster. In the two-site model, we incorporate them via bond vibrations, whereas in the three-site models with the angle fluctuations. Thus, the flexible models have 1 extra DOF compared to the rigid counterparts. However, as can be seen in \cref{flex}, for the three-site models the flexible versions perform worse than the rigid ones. For the two-site model, the flexible model is only slightly better than the rigid model. Nonetheless, as flexible models usually demand smaller integration timesteps and consequently have a higher computational cost the model's performance should be more substantial to justify the extra computational resources.

\subsection*{Accuracy vs. efficiency}

\begin{figure*}[h]
\centering
\includegraphics[width = 0.99\textwidth]{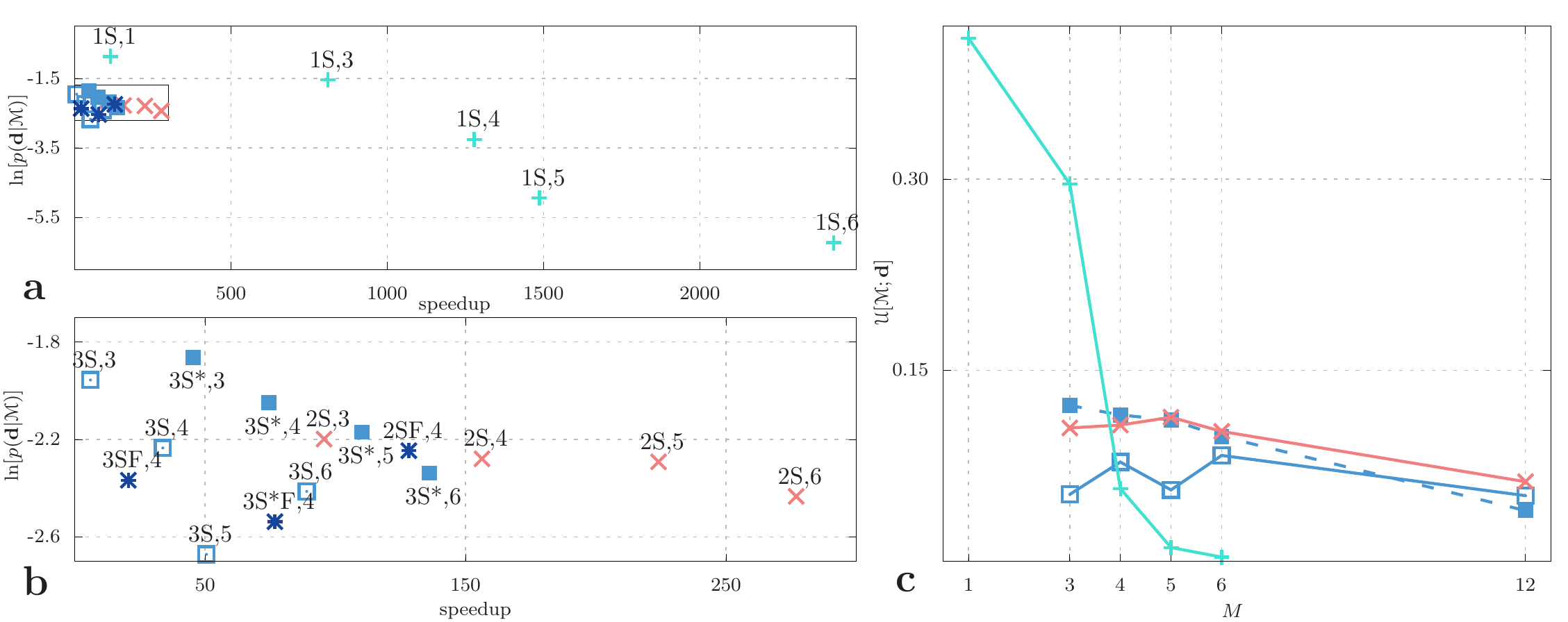}
\caption{Model evidences with respect to the speedup of the examined water models marked with the name and the mapping. The speedup factor is the ratio between the atomistic (TIP4P model) and CG model runtime of $10$~ns, $125$~nm$^3$ {\it NVT} simulation at ambient conditions and maximal integration time step (for CG models see SI, for TIP4P $2$~fs). The boxed section in plot (a) is enlarged in plot (b). {\color{black} Expected utility is shown in (c) for rigid models: {\it 1S} ($+$), {\it 2S} ($\times$), {\it 3S} ($\boxdot$), and {\it 3S}$^*$ ($\blacksquare$) as a function of mapping. }}
\label{logE_CPU}
\end{figure*}

We examine the accuracy vs. efficiency trade-off in \cref{logE_CPU} where we plot the evidence as a function of the speedup compared to the all-atom simulation. 
As a test simulation, we choose the {\it NVE} ensemble simulation at ambient conditions, a cubic domain with an edge of $5$~nm and a simulation length of $10$~ns. We also employ the maximal integration timestep still permitted by the model (see SI). The variation of the runtime varies extensively between the considered models. The computational cost depends on two factors: (i) on the number of particles, which in turn depends on the employed mapping and the number of interaction sites of the model; (ii) on the integration timestep that is increasing with increased coarse-graining since the interactions are softening. For a given mapping, we observe the smallest computational cost for the {\it 1S} model, followed by the {\it 2S}, and {\it 3S*} model while the {\it 3S} model has the highest computational cost. The difference in the {\it 3S} and {\it 3S*} models is due to the smaller timesteps {\color{black} required} by the {\it 3S} model. 

{\color{black} 
The trade-off between the accuracy and efficiency can be formally addressed as a decision problem, where the expected utility \cite{Parmigiani:2010} $\mathcal{U}(\MODEL_i;\DATA)$ of an individual  model $\MODEL_i$  given data $\DATA$ is given by:
\begin{equation}
\mathcal{U}(\MODEL_i;\DATA) = p(\MODEL_i \GIVEN \DATA) u(\MODEL_i). 
\end{equation}
 We define the utility function $u(\MODEL_i)$ as the decimal logarithm of the computational speedup over the atomistic model. As shown in Fig. 6c the model with the maximal expected utility is found for the  {\it 1S} model with $M=1$. When we consider the models incorporating the partial charge: the {\it 3S} model is the most unfavorable, while the {\it 2S} and {\it 3S*} models are comparable in terms of their expected utility. In turn, the appropriate choice for the {\it 2S} and {\it 3S*} models are mappings $M=5$ and $3$, respectively. 
}

\subsection*{Transferability to non-ambient TD conditions}
\begin{figure*}[h]
\centering
\includegraphics[width = 0.7\textwidth]{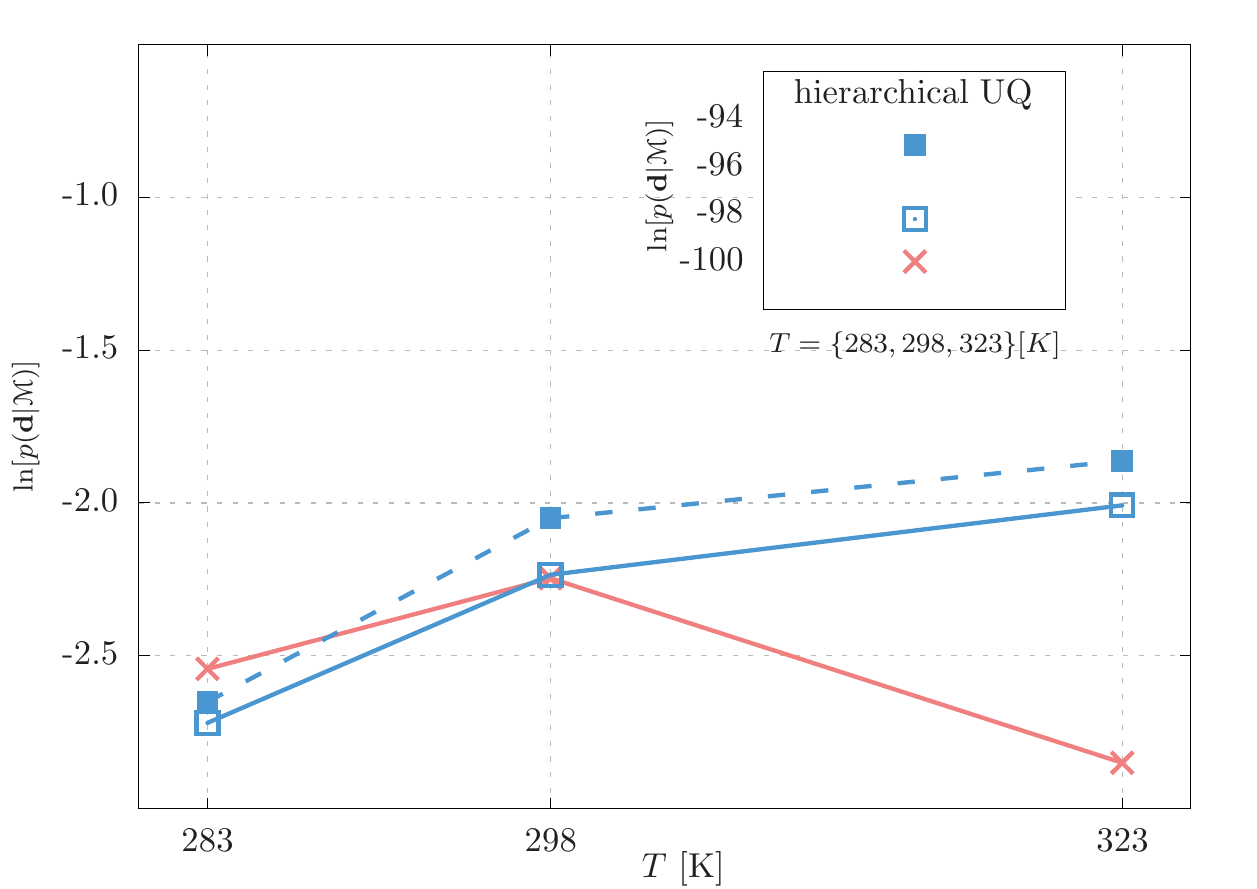}
\caption{Logarithm of the model evidences at different temperatures $T$ for models {\it 2SF} ($\times$), {\it 3S} ($\boxdot$), and {\it 3S*} ($\blacksquare$). The inset shows the model evidences of the hierarchical UQ approach, where all three temperatures are considered concurrently.}
\label{hier}
\end{figure*}

One of the challenges of coarse-graining is the transferability of CG models. Typically, CG models are more sensitive to variations in the thermodynamic conditions than the atomistic models. Furthermore, the more we increase the level of coarse-graining, the more restricted the model is to the thermodynamics state at which it is parametrized. One way of making the model more robust to transferability is to parametrize it for different conditions. Within the Bayesian formalism, the hierarchical UQ allows us to merge multiple QoI. We test the transferability of three models {\it 2SF}, {\it 3S}, and {\it 3S*} for mapping $M=4$. In \cref{hier}, we plot the model evidences for the hierarchical UQ, where the temperatures $T=283, 298, 323$~K are merged and the evidences for the classical UQ at each temperature. 
We observe that the {\it 3S*} model is the most transferable, having the highest hierarchical evidence. For the three-site models, we also observe that it is easier for the CG model to fit higher temperatures. 

\FloatBarrier
\section*{Summary and Discussion}

We propose a data driven, Bayesian framework for the selection of CG water models. We re-examine the CG modeling approach where the mapping and model structure are based on rather ad-hoc assumptions and the system Hamiltonian is derived either by fitting its parameters to relevant experimental data or by deriving the effective interactions from the more detailed, e.g. atomistic simulations. Such a-priori assumptions predefine the accuracy of the model no matter what approach one employs to obtain the Hamiltonian. In this work, we propose a methodology that broadens the investigated space of all possible CG models of liquid water. The Bayesian framework is not constrained to a specific model design but considers many different mappings and model structures. Our model search space encompasses the $1$, $2$, and $3$-site models with either rigid or flexible geometry. We find that for the {\it 1S} model one should consider mappings $M<3$, while for the multiple-site models higher $M$ are more appropriate due to the higher computational cost compared to the single sited models. When choosing between single and multiple-site models one should mainly consider whether the local electrostatics screening is essential for the problem at hand. We observed no significant improvement of models when going from rigid to flexible models, thus implying that one should use rigid geometries for efficiency reasons. The distribution of charge in the three-site models, however, plays an important role as the {\it 3S*} model outperforms the {\it 3S} model and is additionally much cheaper computationally due to the higher maximal integration time step. Additionally, the {\it 3S*} is also the best model in regard to the transferability to non-ambient temperatures. 

{\color{black} 
The methodology presented in this work can be extended to investigate the CG model design of other important chemical and biological systems such as bio-molecules. We emphasize that the data used for the calibration of the models is considered an inherent aspect of the modeling process in a Bayesian framework. The adoption of Bayesian framework in studies of CG models  could quantify the appropriateness of the model designs employed in established CG force fields \cite{Voth:book,Papoian:book} according to target QoIs.
}
The computational limitations associated with Bayesian inference are today largely overcome thanks to the availability of massively parallel computer architectures and a wealth of data is produced by advanced experimental procedures and detailed simulations. This combination enables this 400 year old method  \cite{Stigler:1990, Jaynes:2003} to become a potent alternative to challenging modeling and simulation problems of our times.

\section*{Acknowledgements}
J. Z. acknowledges financial support as an ETH Z\"urich Fellow.
G.A. and P.K. acknowledge support by the European Research Council Advanced Investigator Award 341117.
J.Z. and G.A. acknowledge the help of Lina Kulakova  (ETHZ) in using the software ${\Pi}4U$. The authors thank Matej Praprotnik for useful discussions and critical reading of the manuscript. Finally, the authors acknowledge the computational time at Swiss National Supercomputing Center (CSCS) under the project s659.

\section*{Author contributions statement}
P.K. and J.Z. designed the research, J.Z. and G.A.  conducted the simulations and analysed the results. All authors wrote the manuscript.

\section*{Additional information}

Appendix information contains the experimental data used for target QoI, the properties of the reference water models, the investigated range of the parameters and the maximum a posteriori parameters for all considered models, the integration timesteps used for the computational efficiency evaluation and the details on the hierarchical Bayesian framework.

\subsection*{Competing financial interests}
The authors declare no competing financial interests.

\bibliographystyle{vancouver}
\bibliography{Bibliography}

\begin{thebibliography}{10}

\bibitem{Alberts:1997}
Alberts B, Bray D, Johnson A, Lewis J, Raff M, Roberts K, et~al.
\newblock Essential Cell Biology.
\newblock Garland New York; 1997.

\bibitem{Noid:2013}
Noid WG.
\newblock Perspective: Coarse-grained models for biomolecular systems.
\newblock J Chem Phys. 2013;139:090901.

\bibitem{Shearer2018}
Shearer J, Khalid S.
\newblock Communication between the leaflets of asymmetric membranes revealed
  from coarse-grain molecular dynamics simulations.
\newblock Sci Rep. 2018;8:1805.

\bibitem{Buslaev2017}
Buslaev P, Gushchin I.
\newblock Effects of Coarse Graining and Saturation of Hydrocarbon Chains on
  Structure and Dynamics of Simulated Lipid Molecules.
\newblock Sci Rep. 2017;7:11476.

\bibitem{Bell2017}
Bell DR, Cheng SY, Salazar H, Ren P.
\newblock Capturing RNA Folding Free Energy with Coarse-Grained Molecular
  Dynamics Simulations.
\newblock Sci Rep. 2017;7:45812.

\bibitem{Fajardo2015}
Fajardo OY, Bresme F, Kornyshev AA, Urbakh M.
\newblock Electrotunable Friction with Ionic Liquid Lubricants: How Important
  Is the Molecular Structure of the Ions?
\newblock J Phys Chem Lett. 2015;6:3998--4004.

\bibitem{Riniker:2012}
Riniker S, Allison JR, van Gunsteren W~F.
\newblock On developing coarse-grained models for biomolecular simulation: a
  review.
\newblock Phys Chem Chem Phys. 2012;14:12423--12430.

\bibitem{Foley:2015}
Foley TT, Shell MS, Noid WG.
\newblock The impact of resolution upon entropy and information in
  coarse-grained models.
\newblock J Chem Phys. 2015;143:243104.

\bibitem{Wang:2009}
Wang H, Junghans C, Kremer K.
\newblock Comparative atomistic and coarse-grained study of water: What do we
  lose by coarse-graining?
\newblock Eur Phys J E. 2009;28:221--229.

\bibitem{Drouffe:1991}
Drouffe JM, Maggs AC, Leibler S.
\newblock Computer simulations of self-assembled membranes.
\newblock Science. 1991;254:1353--1356.

\bibitem{Cooke:2005}
Cooke IR, Deserno M.
\newblock Solvent-free model for self-assembling fluid bilayer membranes:
  Stabilization of the fluid phase based on broad attractive tail potentials.
\newblock J Chem Phys. 2005;123:224710.

\bibitem{Shillcock:2005}
Shillcock JC, Lipowsky R.
\newblock Tension-induced fusion of bilayer membranes and vesicles.
\newblock Nat Mater. 2005;4:225 -- 228.

\bibitem{Klein:2001}
Shelley JC, Shelley MY, Reeder RC, Bandyopadhyay S, Klein ML.
\newblock A coarse grained model for phospholipid simulations.
\newblock J Phys Chem B. 2001;105:4464--4470.

\bibitem{Marrink:2004}
Marrink SJ, de~Vries AH, Mark AE.
\newblock Coarse grained model for semiquantitative lipid simulations.
\newblock J Phys Chem B. 2004;108:750--760.

\bibitem{Li:2016}
Li X, Gao L, Fang W.
\newblock Dissipative Particle Dynamics Simulations for Phospholipid Membranes
  Based on a Four-To-One Coarse-Grained Mapping Scheme.
\newblock PLoS ONE. 2016;11:e0154568.

\bibitem{Essex:2011}
Orsi M, Essex JW.
\newblock The ELBA force field for coarse-grain modeling of lipid membranes.
\newblock PLOS Comput Biol. 2011;p. 6:e28637.

\bibitem{Ciccotti2011}
Espa\~nol P, de~la Torre JA, Ferrario M, Ciccotti G.
\newblock Coarse-graining stiff bonds.
\newblock Computational Statistics and Data Analysis. 2011;200:107--129.

\bibitem{Mullinax:2009}
Mullinax JW, Noid WG.
\newblock Extended ensemble approach for deriving transferable coarse-grained
  potentials.
\newblock J Chem Phys. 2009;131:104110.

\bibitem{Das:2012}
Das A, Lu L, Andersen HC, Voth GA.
\newblock The multiscale coarse-graining method. X. Improved algorithms for
  constructing coarse-grained potentials for molecular systems.
\newblock J Chem Phys. 2012;136:194115.

\bibitem{Sinitskiy:2012}
Sinitskiy AV, Saunders MG, Voth GA.
\newblock Optimal Number of Coarse-Grained Sites in Different Components of
  Large Biomolecular Complexes.
\newblock J Phys Chem B. 2012;116:8363--8374.

\bibitem{Arkhipov:2008}
Arkhipov A, Yin Y, Schulten K.
\newblock Four-Scale Description of Membrane Sculpting by BAR Domains.
\newblock Biophys J. 2008;95:2806--2821.

\bibitem{Rudzinski:2014}
Rudzinski JF, Noid WG.
\newblock Investigation of Coarse-Grained Mappings via an Iterative Generalized
  Yvon-Born-Green Method.
\newblock J Phys Chem B. 2014;118:8295--8312.

\bibitem{Zhang:2008}
Zhang Z, Lu L, Noid WG, Krishna V, Pfaendtner J, Voth GA.
\newblock A Systematic Methodology for Defining Coarse-Grained Sites in Large
  Biomolecules.
\newblock Biophys J. 2008;95:5073--5083.

\bibitem{reith}
Reith D, P\"{u}tz M, M\"{u}ller-Plathe F.
\newblock Deriving effective mesoscale potentials from atomistic simulations.
\newblock J Comput Chem. 2003;24:1624--1636.

\bibitem{Voth:2008}
Liu P, Shi Q, Daum{\'e} H, Voth GA.
\newblock A Bayesian statistics approach to multiscale coarse graining.
\newblock J Chem Phys. 2008;129:214114.

\bibitem{Hadley:2012}
Hadley KR, McCabe C.
\newblock Coarse-grained molecular models of water: A review.
\newblock Mol Sim. 2012;38:671--681.

\bibitem{Chaimovich:2009}
Chaimovich A, Shell MS.
\newblock Anomalous waterlike behavior in spherically-symmetric water models
  optimized with the relative entropy.
\newblock Phys Chem Chem Phys. 2009;28:1901--1915.

\bibitem{Izvekov:2005}
Izvekov S, Voth GA.
\newblock Multiscale coarse graining of liquid-state systems.
\newblock J Chem Phys. 2005;123:134105.

\bibitem{Molinero:2009}
Molinero V, Moore EB.
\newblock Water Modeled as an Intermediate Element Between Carbon and Silicon.
\newblock J Phys Chem B. 2009;113:4008--4016.

\bibitem{Jagla:1999}
Jagla EA.
\newblock Core-softened potentials and the anomalous properties of water.
\newblock J Chem Phys. 1999;111:8980--8986.

\bibitem{Hynninen:2012}
Hynninen T, Dias CL, Mkrtchyan A, Heinonen V, Karttunen M, Foster AS, et~al.
\newblock A molecular dynamics implementation of the 3D Mercedes-Benz water
  model.
\newblock Comp Phys Comm. 2012;183:363--369.

\bibitem{Marrink:2013}
Marrink SJ, Tieleman DP.
\newblock Perspective on the MARTINI model.
\newblock Chem Soc Rev. 2013;42:6801--6822.

\bibitem{adaptiveprotein}
Zavadlav J, Melo MN, Marrink SJ, Praprotnik M.
\newblock Adaptive resolution simulation of an atomistic protein in MARTINI
  water.
\newblock J Chem Phys. 2014;140:054114.

\bibitem{pmw}
Yesylevskyy SO, Sch\"{a}fer LV, Sengupta D, Marrink SJ.
\newblock Polarizable water model for the coarse-grained MARTINI force field.
\newblock PLoS Comput Biol. 2010;6:e1000810.

\bibitem{bmw}
Wu Z, Cui Q, Yethiraj A.
\newblock A new coarse-grained model for water: The importance of electrostatic
  interactions.
\newblock J Phys Chem B. 2010;114:10524--10529.

\bibitem{adaptivepolarizable}
Zavadlav J, Melo MN, Marrink SJ, Praprotnik M.
\newblock Adaptive resolution simulation of polarizable supramolecular
  coarse-grained water models.
\newblock J Chem Phys. 2015;142:244118.

\bibitem{Ha-Duong:2009}
Ha-Duong T, Basdevant N, Borgis D.
\newblock A polarizable coarse-grained water model for coarse-grained proteins
  simulations.
\newblock Chem Phys Lett. 2009;469:79--82.

\bibitem{GunsterenFGwater}
Riniker S, van Gunsteren WF.
\newblock A simple, efficient polarizable coarse-grained water model for
  molecular dynamics simulations.
\newblock J Chem Phys. 2011;134:084110.

\bibitem{Hadley:2010}
Hadley KR, McCabe C.
\newblock On the investigation of the coarse-grained models for water:
  Balancing computational efficienncy and the retention of structural
  properties.
\newblock J Phys Chem. 2010;114:4590--4599.

\bibitem{Angelikopoulos:2012}
Angelikopoulos P, Papadimiriou C, Koumoutsakos P.
\newblock Bayesian uncertainty quantification and propagation in molecular
  dynamics simulations: A high performance computing framework.
\newblock J Chem Phys. 2012;137:144103.

\bibitem{Angelikopoulos:2013}
Angelikopoulos P, Papadimiriou C, M E, Koumoutsakos P.
\newblock Data driven, predictive molecular dynamics for nanoscale flow
  simulations under uncertainty.
\newblock J Phys Chem B. 2013;117:14808--14816.

\bibitem{Kulakova:2017}
Kulakova L, Arampatzis G, Angelikopoulos P, Hadjidoukas P, Papadimitriou C,
  Koumoutsakos P.
\newblock Data driven inference for the repulsive exponent of the Lennard-Jones
  potential in molecular dynamics simulations.
\newblock Sci Rep. 2017;7:16576.

\bibitem{Jacobson:2014}
Jacobson LC, Kirby RM, Molinero V.
\newblock How Short Is Too Short for the Interactions of a Water Potential?
  Exploring the Parameter Space of a Coarse-Grained Water Model Using
  Uncertainty Quantification.
\newblock J Phys Chem B. 2014;118:8190--8202.

\bibitem{Rizzi:2013}
Rizzi F, Jones RE, Debusschere BJ, Knio OM.
\newblock Uncertainty quantification in MD simulations of concentration driven
  ionic flow through a silica nanopore. Sensitivity to physical parameters of
  the pore.
\newblock J Chem Phys. 2013;138:194104.

\bibitem{Farrell:2015}
Farrell K, Tinsley~Oden J, Faghihi D.
\newblock A Bayesian framework for adaptive selection, calibration, and
  validation of coarse-grained models of atomistic systems.
\newblock J Comp Phys. 2015;189-208:214114.

\bibitem{Wu:2015}
Wu S, Angelikopoulos P, Papadimiriou C, Moser R, Koumoutsakos P.
\newblock A hierarchical Bayesian framework for force field selection in
  molecular dynamics simulations.
\newblock Phil Trans R Soc A. 2015;374:20150032.

\bibitem{bmw2}
Braun D, Boresch S, Steinhauser O.
\newblock Transport and dielectric properties of water and the influence of
  coarse-graining: Comparing BMW, SPC/E, and TIP3P models.
\newblock J Chem Phys. 2014;140:064107.

\bibitem{CRC:2004}
Lide DR.
\newblock CRC Handbook of Chemistry and Physics.
\newblock CRC Press LLC; 2004.

\bibitem{Kell:1967}
Kell GS.
\newblock Precise Representation of Volume Properties of Water at One
  Atmosphere.
\newblock J Chem Eng Data. 1967;12:66--69.

\bibitem{Moser:2011}
Cheung SH, Oliver TA, Prudencio EE, Prudhomme S, Moser RD.
\newblock Bayesian uncertainty analysis with applications to turbulence
  modeling.
\newblock Reliability Engineering \& System Safety. 2011;96(9):1137 -- 1149.
\newblock Quantification of Margins and Uncertainties.

\bibitem{Beck2004}
Beck JL, Yuen KV.
\newblock {Model selection using response measurements: Bayesian probabilistic
  approach}.
\newblock Journal of Engineering Mechanics. 2004;130(2):192--203.

\bibitem{Knuth2015}
Knuth KH, Habeck M, Malakar NK, Mubeen AM, Placek B.
\newblock Bayesian evidence and model selection.
\newblock Digital Signal Processing. 2015;47:50 -- 67.

\bibitem{Beck2010}
Beck JL.
\newblock {Bayesian system identification based on probability logic}.
\newblock Structural Control and Health Monitoring. 2010;17:825--847.

\bibitem{Parmigiani:2010}
Giovanni~Parmigiani LYTI.
\newblock Decision Theory: Principles and Approaches.
\newblock John Wiley \& Sons, Ltd; 2010.

\bibitem{Voth:book}
Voth GA, editor.
\newblock Coarse-Graining of Condensed Phase and Biomolecular Systems.
\newblock CRC Press; 2009.

\bibitem{Papoian:book}
Papoian GA, editor.
\newblock Coarse-Grained Modeling of Biomolecules.
\newblock CRC Press; 2017.

\bibitem{Stigler:1990}
Stigler SM.
\newblock The History of Statistics The Measurement of Uncertainty before 1900.
\newblock Harvard University Press; 1990.

\bibitem{Jaynes:2003}
Jaynes ET.
\newblock Probability Theory: The Logic of Science.
\newblock Cambridge University Press; 2003.

\bibitem{Plimpton:1995}
Plimpton S.
\newblock Fast Parallel Algorithms for Short-Range Molecular Dynamics.
\newblock J Comput Phys. 1995;117:1--19.

\bibitem{Ching2007}
Ching J, Chen YC.
\newblock {Transitional Markov chain Monte Carlo method for Bayesian model
  updating, model class selection, and model averaging}.
\newblock Journal of engineering mechanics. 2007;133(7):816--832.

\bibitem{Calderhead2009}
Calderhead B, Girolami M.
\newblock Estimating Bayes factors via thermodynamic integration and population
  MCMC.
\newblock Computational Statistics and Data Analysis. 2009;53(12):4028 -- 4045.

\bibitem{Forrester:2008}
Forrester~A KA Sobester~A.
\newblock Engineering Design via Surrogate Modelling: A Practical Guide.
\newblock Wiley; 2008.

\bibitem{Bishop:1996}
Bishop CM.
\newblock Neural Networks for Pattern Recognition.
\newblock Oxford University Press; 1996.

\bibitem{Wu:2016}
Wu S, Angelikopoulos P, Tauriello G, Papadimitriou C, Koumoutsakos P.
\newblock Fusing heterogeneous data for the calibration of molecular dynamics
  force fields using hierarchical {B}ayesian models.
\newblock J Chem Phys. 2016;145:244112.

\end{thebibliography}

\begin{appendices}

\section{Simulation details}\label{simdetails}
In this section, we provide all the relevant information for the MD simulations and the Bayesian inference.

\subsection{Molecular Dynamics}
The MD simulations are performed with LAMMPS package~\cite{Plimpton:1995}. For the calibration, we use a cubic box with periodic boundary conditions and $512$ CG beads, corresponding to $512, 1024$, and $1536$ particles for the $1, 2$, and $3$ sited models. We use the Velocity Verlet integration with the time step of $2$~fs. In rigid models, the bonds/angles are constrained with RATTLE. The cut-off radius for the nonbonded interactions is $3\sigma$, where $\sigma$ is the free parameter of the LJ interaction. The electrostatic interactions beyond the cut-off are corrected with the particle-particle particle-mesh solver scheme. The temperature and pressure of $1$~atm (for the {\it NpT}) are maintained with Nose-Hover with time constants of $\tau_T=0.1$~ps and $\tau_p=1.0$~ps, respectively. Density $\rho$ is evaluated in the {\it NpT} ensemble. For the surface tension $\gamma$ of the water/vapor interface, the water is surrounded by two vacuum regions in the $z$ direction. It is computed in the {\it NVT} ensemble as $\gamma= L_z/2 \langle p_{zz} - p_{||} \rangle$, where where $L_z$ is the box length in the $z$-direction and the $p_z$ and $p_{||}$ denote the perpendicular and the lateral pressure components. The dielectric constant $\varepsilon$ is computed from the fluctuations of the total dipole moment, while the shear viscosity $\eta$ is calculated according to the Green-Kubo formula as $\eta=V/(3k_bT)\int_0^{\infty}\sum_{a<b}p_{ab}(t)p_{ab}(0)dt$, where $V$ is the system volume, $k_B$ is the Boltzmann constant, and $p_{ab}$ are the off-diagonal components of the pressure tensor ($\alpha\beta=xy, xz, yz$) in the {\it NVT} ensemble. The isothermal compressibility $\kappa$ is evaluated under the {\it NpT} conditions as $\kappa=(\langle V^2 \rangle - \langle V \rangle^2)/(k_b T V)$.

 For the Bayesian inference, the protocol for the simulations is the following: a $100$~ps equilibration in the {\it NVT} ensemble followed by $100$~ps equilibration in the {\it NpT} ensemble followed by $1$~ns production run. For the evaluation of $\gamma$, the system is additionally equilibrated for $100$~ps under {\it NVT} conditions. Additional evaluations of each model's properties are performed for the parameters with the maximum likelihood. We first evaluate the maximum integration time-step $\Delta t$ by performing a series of {\it NVE} simulations varying the $\Delta t$ and monitoring the conservation of the total energy. Using the maximum $\Delta t$, approximately  doubled size of the system, $1$~ns equilibration, and $10$~ns production runs we evaluate the properties $\rho, \varepsilon, \gamma, \kappa$ as stated above. 


\subsection{Uncertainty Quantification}
In most practical applications, the posterior distribution 
\begin{equation} \label{eq:posterior}
p( \PM \GIVEN \DATA, \MODEL)  =  \frac{  p(\DATA \GIVEN \PM, \MODEL)   \; p(\PM | \MODEL)  }{   p( \DATA \GIVEN \MODEL )  }  \COMMA
\end{equation}
can not be expressed analytically. 
Moreover, the normalizing constant $p(\DATA | \MODEL)$ is not known since it is given by integration of the denominator in \cref{eq:posterior} over the potentially high dimensional domain of the parameters. 
Hence, we rely on efficient sampling algorithms to identify the posterior distribution. In the present work, we use the Transitional Markov Chain Monte Carlo algorithm (TMCMC) \cite{Ching2007}. 
The algorithm starts by sampling a \emph{population} from the prior distribution, which usually is trivial to be sampled, and through annealing steps provides samples from the posterior distribution. Three major advantages of this algorithm are its ability to sample multimodal distributions, in cases where single chain MCMC methods may fail, the algorithm provides an asymptotically unbiased estimator of the evidence \cite{Calderhead2009} (see SI), and the fact that it can run in parallel. We report that in all simulations the parameter $\beta^2$ was set to $0.04$.

The parallel property of the algorithm is very important in this study since the model evaluation can take on the order of an hour for execution. An efficient implementation of TMCMC can be found in $\Pi4$U\footnote{$\Pi 4$U can be downloaded from here: \url{https://github.com/cselab/pi4u}}, a platform agnostic high performance framework for uncertainty quantification. The sampling algorithm is further accelerated by using surrogate models based on Gaussian processes (see section \ref{sec:surrogates}) leading to a 10-20\% speedup. 

The data on which the Bayesian uncertainty quantification is performed are experimental data for density, dielectric constant (not used in the {\it LJ} model as it is chargeless), surface tension of the water/vapor interface, isothermal compressibility, and shear viscosity (see SI). We want to allow different levels of error for different data thus we choose the covariance matrix $\Sigma$ in the main text to be equal to,
\begin{equation}
\Sigma = \sigma_n^2 \diag \left(  d_1, 4d_2, 8d_3, 8d_4, 8d_5    \right)^2 \PERIOD
\end{equation}
In order to select the best model, we work as follows. First, we perform the inference on all the parameters of each model using all five QoI. Next, we 
fix all parameters except $\sigma$ and $\epsilon$ of the LJ interaction to the values of the maximum a posterior estimation, i.e., the parameter with the highest probability,
\begin{equation}
\PM^\star = \argmax_{\PM} p(\PM \GIVEN \DATA) \PERIOD
\end{equation}
Finally, we infer the remaining parameters of the models excluding the dielectric constant from the QoI. 

Our motivation for this approach is three-fold. First, we remove the dependency of the evidence on the number of model parameters.
 The complexity of the model is then measured through the computational cost of the model. This measure is more appropriate as two models can have the same number of the parameters but could differ in the computational cost. 
 Second, the evidence of the models can be directly compared to the evidence of the LJ model. Finally, having only $3$ parameters reduces the demand on the population size, thus reducing the computational cost of the sampling algorithm.

\subsection{Surrogate models} \label{sec:surrogates}

The computational cost of the Bayesian UQ lies in the computation of the likelihood term in Eq.~\ref{eq:posterior}.
Each likelihood evaluation implies a model evaluation, i.e., a computational demanding simulation using LAMMPS.
The idea of \textit{surrogate models}\cite{Forrester:2008} is to use special interpolation techniques in order to substitute the heavy model evaluations. 

In this work we choose to use Gaussian Processes~\cite{Bishop:1996} (GPs) as a surrogate for the likelihood.
Following the notation in Bishop\cite{Bishop:1996},
given the observed data $\vec{t}=(t_1, \ldots, t_N)$ that correspond to input $\vec{x}=(x_1, \ldots, x_N)$ the conditional distribution $p(t_{N+1} | \vec{t})$ is given by
\begin{equation}
    p(t_{N+1} | \vec{t}) = \NORMAL (  t_{N+1} \, | \, m(x_{N+1}), \sigma^2(x_{N+1}) ) \COMMA
\end{equation}
and
\begin{equation}
    m(x_{N+1}) = \vec{k}^T \vec{C}^{-1}_N \vec{t} \quad\textrm{ and }\quad \sigma^2(x_{N+1}) = c - \vec{k}^T \vec{C}^{-1}_N \vec{k} \COMMA
\end{equation}
where $\vec{k}_i = k(x_i, x_{N+1})$, $i=1,\ldots,N$, $c=k(x_{N+1}, x_{N+1}) + \beta^{-1}$ and the $i,j$ element of the matrix $\vec{C}_N$ is given by $k(x_i, x_j) + \beta^{-1} \delta_{ij}$ for $i,j=1,\ldots,N$. The scalar $\beta$ is the error assumed on the observed  data.
The kernel function $k(x,y)$ models the fact that points which are close in the input space are expected to have more strongly correlated outputs.
In general, the kernel function depends on hyper-parameters which are learned through an optimization process. In this work, we use the squared exponential kernel.

In the context of TMCMC, the training set $\vec{t}_N$ consists of likelihoods values $t_i$ computed on  parameters $x_i$ based only on the exact model evaluation. Based on these values, a GP model is trained and evaluated on a new point $(t_{N+1},x_{N+1})$. If the relative error $\sigma(x_{N+1})/|m(x_{N+1})|$ is less than a threshold then the likelihood value at $x_{N+1}$ is set to $m(x_{N+1})$, otherwise, the model is computed and the likelihood is evaluated. Here, we choose the  threshold to be 0.05.

In order to make more accurate surrogate models, we choose to train a GP model on a subdomain of the finite initial domain centered on $x_{N+1}$. Here, we choose to train on 25\% of the initial domain.

\section{Hierarchical UQ}

We follow the methodology developed in Ref.\cite{Wu:2016}.
We assume that data comes split in $N$ different datasets: $\vvy=\{\vy_1,\ldots, \vy_N\}$.

The likelihood in the probabilistic model $\M_i$ is $\N(\vy_i \,|\, f(\vx_i; \vp_i), \C_i)$.
We assume that the probability of $\vp_i$ depends on a hyper-parameter $\vhp \in \R^{N_\hp}$ and is given by a PDF $\pr{\vp}[\vhp, \M]$, where $\M$ corresponds to the graph describing the relations between $\vhp$, $\vp_i$ and $\vy_i$, see Figure~\ref{fig:dag}.
\begin{figure*}[t]
    \centering
    \includegraphics[width=\textwidth]{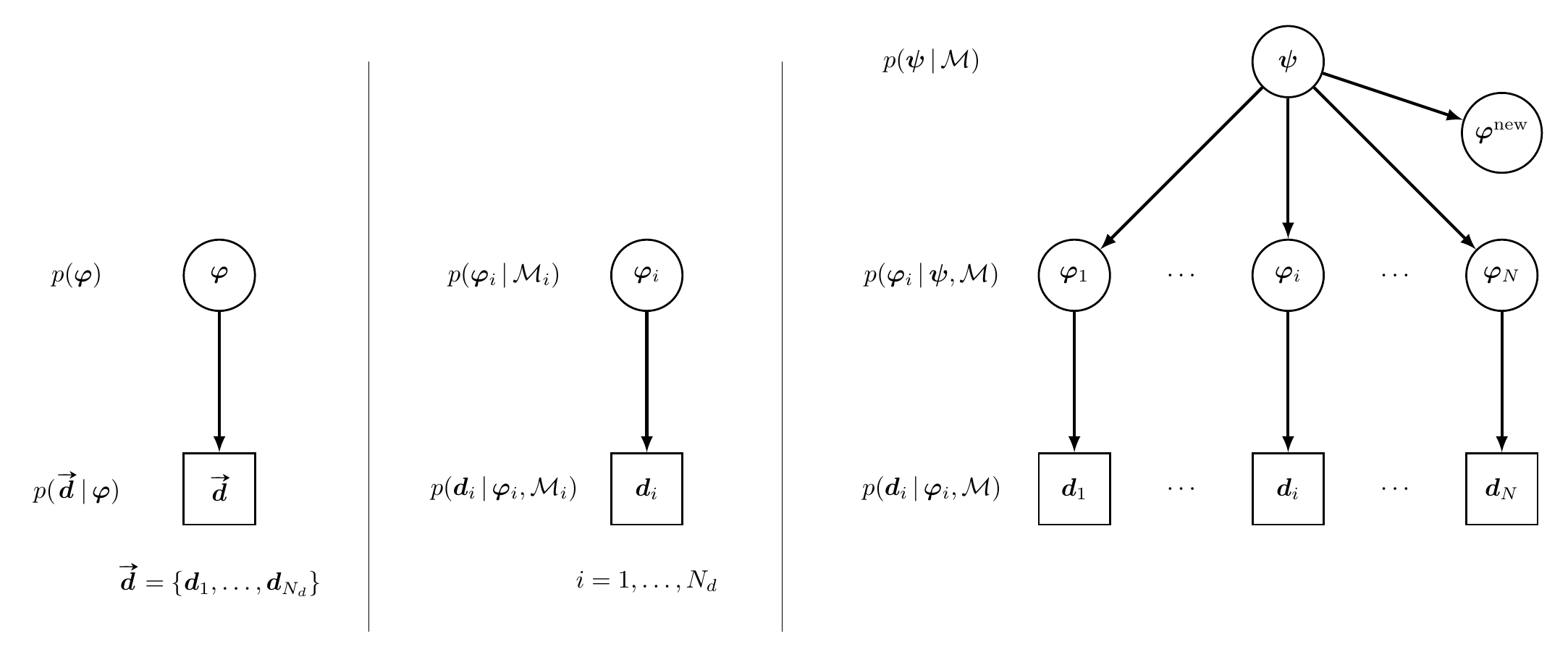}
    \caption{Bayesian networks:  (left) network for the $i$-th dataset, (right) hierarchical Bayesian network.}
    \label{fig:dag}
\end{figure*}
Our goal is to obtain samples from the posterior distribution, $\pr{\vp_i}[\vvy, \M]$,
\begin{equation} \label{eq:h:lik:a}
    \pr{\vp_i}[\vvy, \M] = \int \pr{\vp_i}[\vhp, \vvy, \M] \, 
    \pr{\vhp}[\vvy, \M] \, 
    d\vhp 
    \stp
\end{equation}
 where $\vvy = \{\vy_1, \ldots, \vy_N\}$.
The dependency assumptions from Fig.~\ref{fig:dag} allow to simplify: $\pr{\vp_i}[\vhp, \vvy, \M] = \pr{\vp_i}[\vhp, \vy_i, \M]$, and equation~\eqref{eq:h:lik:a} can be rewritten using the Bayes' theorem:
\begin{equation} \label{eq:h:lik:b}
    \pr{\vp_i}[\vvy, \M] = 
    \int \frac{ \pr{\vy_i}[\vp_i, \vhp, \M] \, \pr{\vp_i}[\vhp, \M] }{ \pr{\vy_i}[\vhp, \M] } \, \pr{\vhp}[\vvy, \M] \, d\vhp \stp
\end{equation}
Since $\pr{\vy_i}[\vp_i, \vhp, \M] = \pr{\vy_i}[\vp_i, \M]$, equation~\eqref{eq:h:lik:b} simplifies to
\begin{equation}
    \pr{\vp_i}[\vvy, \M] = 
      \quad \pr{\vy_i}[\vp_i, \M] \int \frac{ \pr{\vp_i}[\vhp, \M] }{ \pr{\vy_i}[\vhp, \M] } \, \pr{\vhp}[\vvy, \M] \, d\vhp \stp
\end{equation}
Finally, the posterior distribution (\ref{eq:h:lik:a}) can be approximated as
\begin{equation}
    \pr{\vp_i}[\vvy, \M] \approx \frac{ \pr{\vy_i}[\vp_i, \M] }{ N_s } \sum_{k=1}^{N_s} \frac{ \pr{\vp_i}[\vhp^{(k)}, \M] }{ \pr{\vy_i}[\vhp^{(k)}, \M] } \cnt
\end{equation}
where, $\vhp^{(k)}\sim \pr{\vhp}[\vvy, \M]$ and $N_s$ is sufficiently large.
Thus, in order to obtain $\vp_i$ samples, we first have to sample the probability distribution $\pr{\vhp}[\vvy, \M]$, which, according to Bayes' theorem, is equal to
\begin{equation}
    \pr{\vhp}[\vvy, \M] = \frac{ \pr{\vvy}[\vhp, \M] \, \pr{\vhp}[\M] }{ \pr{\vvy}[\M] } \cnt
\end{equation}
where $\pr{\vhp}[\M]$ is the prior PDF on $\vhp$ and $\pr{\vvy}[\M]$ is the normalizing constant.
Exploiting the dependency assumption of Fig.~\ref{fig:dag} we see that
\begin{equation}
    \pr{\vvy}[\vhp, \M] = \prod_{i=1}^{N} \pr{\vy_i}[\vhp, \M] \cnt
\end{equation}
and the likelihood of $i$-th dataset can be expressed according to the total probability theorem as
\begin{equation} \label{eq:h:lik:c}
    \pr{\vy_i}[\vhp, \M] = \int \pr{\vy_i}[\vp_i, \M] \, \pr{\vp_i}[\vhp, \M] \, d\vp_i \stp
\end{equation}
Here we introduce the model $\M_i$ described in Fig.~\ref{fig:dag}.
The posterior distribution of this model will be used as instrumental density for important sampling.
Under the modeling assumption $\pr{\vy_i}[\vp_i, \M] = \pr{\vy_i}[\vp_i, \M_i]$ (see \cite{Wu:2016}) and the use of the Bayes' theorem, equation~\eqref{eq:h:lik:c} is written as
\begin{equation}
       \pr{\vy_i}[\vhp, \M] =
       \int \frac{ \pr{\vp_i}[\vy_i, \M_i] \, \pr{\vy_i}[\M_i] }{ \pr{\vp_i}[\M_i] } \, \pr{\vp_i}[\vhp, \M] \, d\vp_i \cnt
\end{equation}
or, equivalently, as
\begin{equation}
    \pr{\vy_i}[\vhp, \M] =
    \pr{\vy_i}[\M_i] \int \frac{ \pr{\vp_i}[\vhp, \M] }{ \pr{\vp_i}[\M_i] } \, \pr{\vp_i}[\vy_i, \M_i] \, d\vp_i \stp
\end{equation}
Finally, equation~\eqref{eq:h:lik:c} can be approximated as
\begin{equation}
    \pr{\vy_i}[\vhp, \M] \approx
    \frac{ \pr{\vy_i}[\M_i] }{ N_s } \sum_{k=1}^{N_s} \frac{ \pr{\vp_i^{(k)}}[\vhp, \M] }{ \pr{\vp_i^{(k)}}[\M_i] } \cnt
\end{equation}
where $\vp_i^{(k)} \sim \pr{\vp_i}[\vy_i, \M_i]$ and $N_s$ is sufficiently large.
Note that in general $N_s$ can be different for each data set $\vy_i$.
The advantage of this approach is that the likelihoods $\pr{\vy_i}[\vp_i, \M_i],\, i=1,\ldots,N$,  which are the most expensive part of the computations, are not re-evaluated for each $\vhp$.

\section{UQ inference for density, dielectric constant, surface tension, isothermal compressibility, and shear viscosity}

A uniform prior distribution is used for all model parameters. The population size of the TMCMC is 3000 for the {\it 2S} model. For the {\it 2SF} {\it 3S}, and {\it 3S*} model it is 4000, while for the {\it 3SF} and {\it 3S*F} models it is 5000.

\begin{table}[H]
\begin{flushleft}
    \begin{tabular}{ c | c c c c c c }
    {\it 2S} \\ 
    $M$ & $\sigma$ & $\epsilon$ & $e$ & $l$ & $\sigma$ & $\mu$  \\ \hline
    3 & 4.15 - 4.55 & 0.30 - 1.00 & 0.30 - 1.01 & 1.1 - 2.8 & 0.0 - 0.1 & 4.4 - 5.9  \\
    4 & 4.60 - 5.00 & 0.30 - 1.00 & 0.30 - 1.01 & 1.1 - 2.8 & 0.0 - 0.1 & 5.1 - 6.6  \\
    5 & 5.00 - 5.40 & 0.30 - 1.00 & 0.30 - 1.01 & 1.1 - 2.8 & 0.0 - 0.1 & 5.4 - 6.9  \\
    6 & 5.30 - 5.70 & 0.30 - 1.00 & 0.30 - 1.01 & 1.1 - 2.8 & 0.0 - 0.1 & 5.8 - 7.3  \\
  12 & 6.50 - 7.10 & 0.30 - 1.15 & 0.30 - 1.01 & 1.7 - 3.2 & 0.0 - 0.1 & 8.0 - 11.0  \\
  \end{tabular}
        \begin{tabular}{ c | c c c c c c c}
      {\it 3S} \\ 
    $M$ & $\sigma$& $\epsilon$& $e$ & $l$ & $\theta$ & $\sigma$ & $\mu$ \\ \hline
    3 & 4.15 - 4.55 & 0.30 - 1.00 & 0.30 - 1.01 & 1.1 - 2.8 & 0 - 140 & 0.0 - 0.1  & 4.4 - 5.9 \\
    4 & 4.60 - 5.00 & 0.30 - 1.00 & 0.30 - 1.01 & 1.1 - 2.8 & 0 - 140 & 0.0 - 0.1  & 5.1 - 6.6 \\
    5 & 5.00 - 5.40 & 0.30 - 1.00 & 0.30 - 1.01 & 1.1 - 2.8 & 0 - 140 & 0.0 - 0.1  & 5.4 - 6.9 \\
    6 & 5.30 - 5.70 & 0.30 - 1.00 & 0.30 - 1.01 & 1.1 - 2.8 & 0 - 140 & 0.0 - 0.1 & 5.8 - 7.3  \\ 
  12 & 6.50 - 7.10 & 0.30 - 1.15 & 0.30 - 1.01 & 1.7 - 3.2 & 0 - 140 & 0.0 - 0.1 & 8.0 - 11.0\\ 
  \end{tabular}
          \begin{tabular}{ c | c c c c c c c}
      {\it 3S*} \\ 
    $M$ & $\sigma$ & $\epsilon$ & $e$ & $l$ & $\theta$ & $\sigma$ & $\mu$ \\ \hline
    3 & 4.15 - 4.55 & 0.30 - 1.00 & 0.30 - 1.01 & 0.8 - 1.5 & 40 - 180 & 0.0 - 0.1  & 4.4 - 5.9 \\
    4 & 4.60 - 5.00 & 0.30 - 1.00 & 0.30 - 1.01 & 0.8 - 1.5 & 40 - 180 & 0.0 - 0.1  & 5.1 - 6.6 \\
    5 & 5.00 - 5.40 & 0.30 - 1.00 & 0.30 - 1.01 & 0.8 - 1.5 & 40 - 180 & 0.0 - 0.1  & 5.4 - 6.9 \\
    6 & 5.30 - 5.70 & 0.30 - 1.00 & 0.30 - 1.01 & 0.8 - 1.5 & 40 - 180 & 0.0 - 0.1  & 5.8 - 7.3  \\ 
  12 & 6.70 - 7.30 & 0.30 - 1.15 & 0.30 - 1.01 & 1.2 - 2.0 & 40 - 180 & 0.0 - 0.1  & 8.0 - 11.0\\ 
  \end{tabular}
\end{flushleft}
\caption{Model parameter ranges ($\sigma$ in \AA, $\epsilon$ in kcal/mol, $e$ in $e_0$,  $l$ in \AA, and $\theta$ in $^{\circ}$) for the considered rigid models and mappings $M$. The dipole moment $\mu$ is given in D.}
\label{range}
\end{table}

\begin{table}[H]
\begin{flushleft}
        \begin{tabular}{ c | c c c c c c}
    model & $\sigma$ & $\epsilon$ & $e$ & $l$ & $k_B$ & $\mu$ \\ \hline
    {\it 2SF} & 4.60 - 5.00 & 0.30 - 1.00 & 0.30 - 1.01 & 1.1 - 2.8  & 10 - 60 & 5.1 - 6.6\\
  \end{tabular}
          \begin{tabular}{ c | c c c c c c c c }
    model & $\sigma$ & $\epsilon$ & $e$ & $l$ & $\theta$ & $k_A$ & $\sigma$ & $\mu$ \\ \hline
    {\it 3SF} & 4.60 - 5.00 & 0.30 - 1.00 & 0.30 - 1.01 & 1.1 - 2.8 & 0 - 140 & 1 - 10 & 0.0 - 0.1 & 5.1 - 6.6 \\
  \end{tabular}
            \begin{tabular}{ c | c c c c c c c c}
    model & $\sigma$ & $\epsilon$ & $e$ & $l$  & $\theta$ & $k_A$ & $\sigma$ & $\mu$ \\ \hline
   {\it 3S*F} & 4.60 - 5.00 & 0.30 - 1.00 & 0.30 - 1.01 & 0.8 - 1.5 & 40 - 180 & 1 - 10 & 0.0 - 0.1 & 5.1 - 6.6 \\
  \end{tabular}
\end{flushleft}
\caption{Model parameter ranges ($\sigma$ in \AA, $\epsilon$ in kcal/mol, $e$ in $e_0$,  $l$ in \AA, $k_B$ in kcal/mol/nm$^2$, $\theta$ in $^{\circ}$, and $k_A$ in kcal/mol) for the considered flexible models and mappings $M=4$. The dipole moment $\mu$ is given in D.}
\label{range}
\end{table}

\begin{table}[H]
\begin{flushleft}
    \begin{tabular}{ c | c c c c c }
    {\it 2S} \\ 
    $M$ & $\sigma$ & $\epsilon$ & $e$ & $l$ & $\mu$   \\ \hline
    3 & 4.3912 & 0.9699 & 0.6213 & 1.5056 & 4.4934 \\ 
    4 & 4.8491 & 0.9690 & 0.6596 & 1.6880 & 5.3478 \\
    5 & 5.2222 & 0.9982 & 0.7197 & 1.6171 & 5.5908 \\
    6 & 5.5018 & 0.9928 & 0.8683 & 1.3967 & 5.8255 \\ 
  12 & 6.9531 & 0.9662 & 0.7707 & 2.5069 & 9.2808 \\ 
  \end{tabular}
        \begin{tabular}{ c | c c c c c c}
      {\it 3S} \\ 
    $M$ & $\sigma$ & $\epsilon$ & $e$ & $l$ & $\theta$ & $\mu$ \\ \hline
    3 & 4.4119 & 0.9512 & 0.6047 & 1.6138 & 3.5880 & 4.6850    \\
    4 & 4.8468 & 0.8782 & 0.6285 & 1.8062 & 7.1728 & 5.4420   \\
    5 & 5.2159 & 0.9926 & 0.7262 & 1.6778 & 6.9914 & 5.8413   \\
    6 & 5.5299 & 0.9808 & 0.6428 & 2.1903 & 9.0216 & 6.7420   \\ 
  12 & 6.9527 & 1.0053 & 0.8351 & 2.3593 & 53.9798 & 8.4340   \\ 
  \end{tabular}
          \begin{tabular}{ c | c c c c c c}
      {\it 3S*} \\ 
    $M$ & $\sigma$ & $\epsilon$ & $e$ & $l$ & $\theta$ & $\mu$ \\ \hline
    3 & 4.4040 & 0.8625 & 1.0053 & 0.9551 & 57.4459 & 4.4310    \\
    4 & 4.8898 & 0.9872 & 0.8219 & 1.2458 & 65.4964 & 5.3182   \\
    5 & 5.2212 & 0.9654 & 0.9112 & 1.1455 & 66.0899 & 5.4656 \\
    6 & 5.5341 & 0.9943 & 0.9143 & 1.2926 & 62.9153 & 5.9224 \\ 
  12 & 7.0897 & 0.9186 & 0.8780 & 1.8466 & 86.7040 & 10.6878 \\ 
  \end{tabular}
\end{flushleft}
\caption{Model parameters ($\sigma$ in \AA, $\epsilon$ in kcal/mol, $e$ in $e_0$,  $l$ in \AA, $k_B$ in kcal/mol/nm$^2$, $\theta$ in $^{\circ}$, and $k_A$ in kcal/mol) with maximum likelihood for the considered rigid models and mappings $M$. The dipole moment $\mu$ is given in D.}
\label{best}
\end{table}

\begin{table}[H]
\begin{flushleft}
        \begin{tabular}{ c | c c c c c c}
    model & $\sigma$ & $\epsilon$ & $e$ & $l$& $k_B$ & $\mu$ \\ \hline 
    {\it 2SF} & 4.9172 & 0.9396 & 0.5452 & 2.4337 & 34.8621 & 6.3730  \\      
  \end{tabular}
            \begin{tabular}{ c | c c c c c c c}
    model & $\sigma$ & $\epsilon$ & $e$ & $l$ & $\theta$ & $k_A$ & $\mu$  \\ \hline
    {\it 3SF} & 4.8348 & 0.9395 & 0.6493 & 1.7219 & 29.3119 & 9.2364 & 5.1953 \\
  \end{tabular}
          \begin{tabular}{ c | c c c c c c c}
    model & $\sigma$ & $\epsilon$ & $e$ & $l$ & $\theta$ & $k_A$ & $\mu$ \\ \hline
    {\it 3S*F} & 4.8982 & 0.9092 & 0.5495 & 1.4730 & 87.0474 & 7.8318 & 5.3524  \\
  \end{tabular}
\end{flushleft}
\caption{Model parameters ($\sigma$ in \AA, $\epsilon$ in kcal/mol, $e$ in $e_0$,  $l$ in \AA, $k_B$ in kcal/mol/nm$^2$, $\theta$ in $^{\circ}$, and $k_A$ in kcal/mol) with maximum likelihood for the considered flexible models and mappings $M=4$. The dipole moment $\mu$ is given in D.}
\label{best}
\end{table}

\section{UQ inference for density, surface tension, isothermal compressibility, and shear viscosity}

A uniform prior distribution is used for all model parameters. The population size of the TMCMC is 2000 for all models.

\begin{table}[H]
\begin{flushleft}
  \begin{tabular}{ c | c c c  }
    {\it LJ} \\ 
    $M$ & $\sigma$  & $\epsilon$ & $\sigma$  \\ \hline
    1 & 2.80 - 3.00 & 0.50 - 1.25 & 0.0 - 0.1 \\
    3 & 4.00 - 4.40 & 0.50 - 1.25 & 0.0 - 0.1  \\
    4 & 4.45 - 4.85 & 0.50 - 1.25 & 0.0 - 0.1 \\
    5 & 4.80 - 5.30 & 0.50 - 1.25 & 0.0 - 0.1  \\
    6 & 5.10 - 5.60 & 0.50 - 1.25 & 0.0 - 0.1 \\ 
  \end{tabular}
    \begin{tabular}{ c | c c c | c c c }
     {\it 2S} & & & & & fixed\\ 
   $M$ & $\sigma$ & $\epsilon$ & $\sigma$ & $e$ & $l$  & $\mu$  \\ \hline
    3 & 4.15 - 4.55 & 0.30 - 1.15 & 0.0 - 0.1 & 0.6213 & 1.5056 & 4.4934  \\
    4 & 4.60 - 5.00 & 0.30 - 1.15 & 0.0 - 0.1 & 0.6596 & 1.6880 & 5.3478  \\
    5 & 5.00 - 5.40 & 0.30 - 1.15 & 0.0 - 0.1 & 0.7197 & 1.6171 & 5.5908  \\
    6 & 5.30 - 5.70 & 0.30 - 1.15 & 0.0 - 0.1 & 0.8683 & 1.3967 & 5.8255  \\
  12 & 6.50 - 7.10 & 0.30 - 1.15 & 0.0 - 0.1 & 0.7707 & 2.5069 & 9.2808  \\
  \end{tabular}
        \begin{tabular}{ c | c c c | c c c c}
      {\it 3S} & & & & & fixed\\   
    $M$ & $\sigma$& $\epsilon$ & $\sigma$ & $e$ & $l$ & $\theta$ & $\mu$ \\ \hline
    3 & 4.15 - 4.55 & 0.30 - 1.15 & 0.0 - 0.1 & 0.6047 & 1.6138 & 3.5880 & 4.685 \\
    4 & 4.60 - 5.00 & 0.30 - 1.15 & 0.0 - 0.1& 0.6285 & 1.8062 & 7.1728 & 5.4420 \\
    5 & 5.00 - 5.40 & 0.30 - 1.15 & 0.0 - 0.1 & 0.7262 & 1.6778 & 6.9914 & 5.8413 \\
    6 & 5.30 - 5.70 & 0.30 - 1.15 & 0.0 - 0.1 & 0.6428 & 2.1903 & 9.0216 & 6.7420  \\ 
  12 & 6.50 - 7.10 & 0.30 - 1.15 & 0.0 - 0.1 & 0.8351 & 2.3593 & 53.9798 & 8.4340 \\ 
  \end{tabular}
          \begin{tabular}{ c | c c c | c c c c}
      {\it 3S*} & & & & & fixed \\ 
    $M$ & $\sigma$ & $\epsilon$ & $\sigma$ & $e$ & $l$ & $\theta$ & $\mu$ \\ \hline
    3 & 4.15 - 4.55 & 0.30 - 1.15 & 0.0 - 0.1 & 1.0053 & 0.9551 & 57.4459 & 4.4310 \\
    4 & 4.60 - 5.00 & 0.30 - 1.15 & 0.0 - 0.1 & 0.8219 & 1.2458 & 65.4964 & 5.3182 \\
    5 & 5.00 - 5.40 & 0.30 - 1.15 & 0.0 - 0.1 & 0.9112 & 1.1455 & 66.0899 & 5.4656 \\
    6 & 5.30 - 5.70 & 0.30 - 1.15 & 0.0 - 0.1 & 0.9143 & 1.2926 & 62.9153 & 5.9224  \\ 
  12 & 6.70 - 7.30 & 0.30 - 1.15 & 0.0 - 0.1 & 0.8780 & 1.8466 & 86.7040 & 10.6878\\ 
  \end{tabular}
\end{flushleft}
\caption{Model parameter ranges ($\sigma$ in \AA, $\epsilon$ in kcal/mol, $e$ in $e_0$,  $l$ in \AA, and $\theta$ in $^{\circ}$) for the considered rigid models and mappings $M$. The dipole moment $\mu$ is given in D.}
\label{range}
\end{table}

\begin{table}[H]
\begin{flushleft}
        \begin{tabular}{ c | c c c | c c c c }
      & & & & & fixed  \\ 
   model & $\sigma$ & $\epsilon$ & $\sigma$ & $e$ & $l$& $k_B$ &  $\mu$ \\ \hline
    {\it 2SF} & 4.60 - 5.00 & 0.30 - 1.15 & 0.0 - 0.1 & 0.5452 & 2.4337 & 34.8621 & 6.3730   \\
  \end{tabular}
        \begin{tabular}{ c | c c c | c c c c c }
      & & & & & fixed  \\ 
   model & $\sigma$ & $\epsilon$ & $\sigma$ & $e$ & $l$& $\theta$ & $k_A$ & $\mu$ \\ \hline
    {\it 3SF} & 4.60 - 5.00 & 0.30 - 1.15 & 0.0 - 0.1 & 0.6493 & 1.7219 & 29.3119 & 9.2364 & 5.1953 \\
  \end{tabular}
        \begin{tabular}{ c | c c c | c c c c c }
      & & & & & fixed  \\ 
   model & $\sigma$ & $\epsilon$ & $\sigma$ & $e$ & $l$& $\theta$ & $k_A$ & $\mu$ \\ \hline
   {\it 3S*F} & 4.60 - 5.00 & 0.30 - 1.15 & 0.0 - 0.1 & 0.5495 & 1.4730 & 87.0474 & 7.8318 & 5.3524 \\
  \end{tabular}
\end{flushleft}
\caption{Model parameter ranges ($\sigma$ in \AA, $\epsilon$ in kcal/mol, $e$ in $e_0$,  $l$ in \AA, $k_B$ in kcal/mol/nm$^2$, $\theta$ in $^{\circ}$, and $k_A$ in kcal/mol) for the considered flexible models and mapping $M=4$. The dipole moment $\mu$ is given in D.}
\label{range}
\end{table}

\begin{table}[H]
\begin{flushleft}
  \begin{tabular}{ c | c c  }
   &  {\it LJ} \\ 
    $M$ & $\sigma$ & $\epsilon$ \\ \hline
    1 & 2.9534 & 1.0140 \\
    3 & 4.3353 & 1.1826  \\
    4 & 4.7360 & 1.2094  \\
    5 & 5.0505 & 1.2369  \\
    6 & 5.3938 & 1.2056 \\ 
  \end{tabular}
   \\ \begin{tabular}{ c | c c |  c c |  c c}
     & {\it 2S} & &  3S & & 3S* \\ 
    $M$ & $\sigma$ & $\epsilon$  & $\sigma$ & $\epsilon$ & $\sigma$ & $\epsilon$  \\ \hline
    3 & 4.4079 & 0.9329  & 4.3953 & 1.0089 & 4.4309 & 0.9398\\ 
    4 & 4.8604 & 0.9522  & 4.8374 & 1.0255 & 4.8759 & 0.9128\\
    5 & 5.2117 & 0.9996  & 5.2111 & 1.0244  & 5.2280 & 1.0223 \\
    6 & 5.5137 & 1.0496  & 5.5475 & 0.9468  & 5.5700 & 1.0464\\ 
  12 & 7.0086 & 1.0068  & 6.9733 & 1.0452 & 7.1020 & 0.9948 \\ 
  \end{tabular}
    \\  \begin{tabular}{ c | c c |  c c |  c c}
   &  {\it 2SF} & &  3SF & & 3S*F\\ 
    $M$ & $\sigma$ & $\epsilon$ & $\sigma$ & $\epsilon$ & $\sigma$ & $\epsilon$ \\ \hline 
    4 & 4.8967 & 0.9923 & 4.8328 & 1.0120 & 4.9210 & 0.8871 \\      
      \end{tabular}
\end{flushleft}
\caption{Model parameters ($\sigma$ in \AA, $\epsilon$ in kcal/mol) with maximum likelihood for the considered models and mappings $M$.}
\label{best}
\end{table}

\section{Maximum integration timestep}
\begin{table}[H]
\begin{center}
  \begin{tabular}{ c | c c c c c c c }
    $M$/model & {\it LJ} & {\it 2S} & {\it 2SF} &  {\it 3S} & {\it 3SF} &  {\it 3S*} & {\it 3S*F} \\ \hline \hline
    1 & 10 & /    & /    & /    & / & /    & / \\
    3 & 26 & 22 & /    & 2   & / & 20 & / \\
    4 & 34 & 28 & 22 & 8   &6 & 26 & 28 \\
    5 & 38 & 34 & /    &12  & / & 32 & /\\
    6 & 48 & 40 & /    &18  & / & 36 & /\\ 
   12 & /   & 70 & /    & 66 & / & 66 & /\\ 
  \end{tabular}
\end{center}
\caption{Maximum integration timestep (in fs) for the considered models and mappings $M$.}
\label{maxT}
\end{table}

\section{Information about hyper-parameter models for hierarchical inference} \label{sec:hb_info}

\renewcommand{\arraystretch}{1.2}
\begin{table}[h!]
        \caption{Parameters of the prior for the hyper-parameters for the uniform and the gamma prior models.}
        \label{table:hyperparameters}
        \centering
        \begin{tabular}{ c | c | c }
            & $ \M=\MU$ & $\M=\Gamma$ \\  \hline
            $[a_{1}^{\M}, b_{1}^{\M}]$ & [0.0, 8.0]  & [3.0, 6.0] \\ \hline
            $[a_{2}^{\M}, b_{2}^{\M}]$ & [0.0, 5.0]  & [0.0, 1.0] \\ \hline
            $[a_{3}^{\M}, b_{3}^{\M}]$ & [0.0, 5.0]  & [0.1, 5.0] \\ \hline
            $[a_{4}^{\M}, b_{4}^{\M}]$ & [0.0, 5.0]  & [0.0, 1.0] \\ \hline
            $[a_{5}^{\M}, b_{5}^{\M}]$ & [0.0, 5.0]  & [0.1, 2.0] \\ \hline
            $[a_{6}^{\M}, b_{6}^{\M}]$ & [0.0, 1.0]  & [0.0, 3.0] \\
        \end{tabular}
\end{table}
\renewcommand{\arraystretch}{1}

We assume that,
\begin{equation}
    \pr{\vp}[\vhp, \M] = \prod_{j=1}^3 \pr{\p_j}[\vhp, \M] \cnt
\end{equation}
and we consider the following two models: 
\begin{itemize}
    \item[1)] uniform: $\pr{\p_j}[\vhp, \M] = \U( \p_j \,|\, \hp_{2j-1}, \hp_{2j-1} + \hp_{2j})$, where $\U(\cdot | a, b)$ is the uniform distribution with parameters $a, b$ and $\M$ is set to $\MU$,
    \item[2)]  gamma: $\pr{\p_j}[\vhp, \M] = \Gamma(\p_j \,|\, \hp_{2j-1}, \hp_{2j})$, where $\Gamma(\cdot | \mu, \sigma)$ is a re-parametrized gamma distribution with mean $\mu$ and standard deviation $\sigma$ and $\M$ is set to $\Gamma$. Given $\mu$ and $\sigma$ the shape and the scale of the gamma distribution are given by $k=\frac{\mu^2}{\sigma}$ and $\theta=\frac{\sigma}{\mu}$.
\end{itemize}
The prior distribution on the hyper-parameters is modeled as independent uniform,
\begin{equation} \label{eq:hp_12}
    \pr{\vhp}[\M] = \prod_{j=1}^6 \U( \hp_j \,|\, a_j^{\M}, b_j^{\M}) \cnt
\end{equation}
where $\M \in \{\MU,\Gamma \}$ and the constants $a_j^{\M}, b_j^{\M}$ are given in Table~\ref{table:hyperparameters}. The model $\U$ is according to the Bayesian model selection criterion, an order of magnitude more plausible and thus will be used for the further inference.

\begin{figure*}[t]
    \centering
    \includegraphics[width=0.5\textwidth]{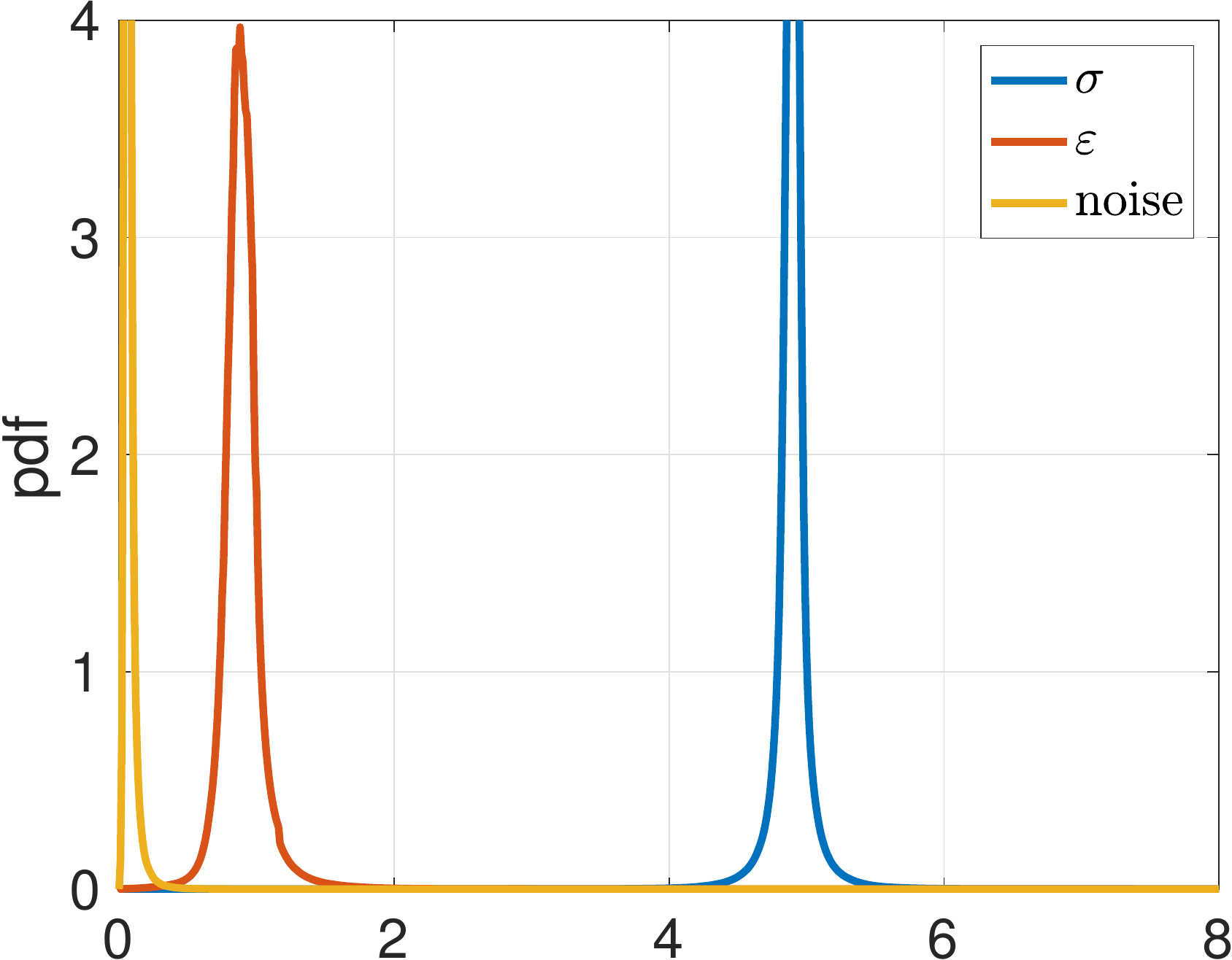}
    \caption{$\pr{\vp^{\textrm{new}}}[ \protect\vvy, \U]$.}
    \label{fig:new:param}
\end{figure*}

\hspace{50pt}
\renewcommand{\arraystretch}{1.2}
\begin{table}[h!]
        \caption{Logarithm of the evidence for the uniform and the gamma prior models.}
        \label{table:hierLogE}
        \centering
        \begin{tabular}{ c | c | c }
            & $ \M=\MU$ & $\M=\Gamma$ \\  \hline
{\it 3S2 } &  -95.1113   &   -98.2459 \\ \hline
{\it 3S }  &  -98.1956  &   -98.6998  \\ \hline
{\it 2SF } &  -99.9770  &   -99.1716 \\ 
        \end{tabular}
\end{table}
\renewcommand{\arraystretch}{1}

\end{appendices}

\end{document}